\documentclass[]{elsarticle} 

\usepackage[hyphens]{url}

  \journal{arXiv.org} 

\usepackage{lineno} 

\usepackage{graphicx}

\usepackage[T1]{fontenc}
\usepackage{lmodern}
\usepackage{amssymb,amsmath}
\usepackage{ifxetex,ifluatex}
\usepackage{fixltx2e} 
\IfFileExists{upquote.sty}{\usepackage{upquote}}{}
\ifnum 0\ifxetex 1\fi\ifluatex 1\fi=0 
  \usepackage[utf8]{inputenc}
\else 
  \usepackage{fontspec}
  \ifxetex
    \usepackage{xltxtra,xunicode}
  \fi
  \defaultfontfeatures{Mapping=tex-text,Scale=MatchLowercase}
  
\fi
\IfFileExists{microtype.sty}{\usepackage{microtype}}{}
\usepackage[margin=1.0in]{geometry}
\usepackage[]{natbib}

\ifxetex
  \usepackage[setpagesize=false, 
              unicode=false, 
              xetex]{hyperref}
\else
  \usepackage[unicode=true]{hyperref}
\fi
\hypersetup{breaklinks=true,
            bookmarks=true,
            pdfauthor={},
            pdftitle={Individualized treatment effect was predicted best by modeling baseline risk in interaction with treatment assignment},
            colorlinks=false,
            urlcolor=blue,
            linkcolor=magenta,
            pdfborder={0 0 0}}

\newlength{\cslhangindent}
\newlength{\csllabelwidth}
\newlength{\cslentryspacingunit} 
  {}%
  {\par}
\newenvironment{CSLReferences}[2] 
 {
  \setlength{\parindent}{0pt}
  \ifodd #1
  \let\oldpar\par
  \def\par{\hangindent=\cslhangindent\oldpar}
  \fi
  \setlength{\parskip}{#2\cslentryspacingunit}
 }%
 {}
\usepackage{calc}

\usepackage{setspace}
\usepackage{amsmath}
\usepackage{amssymb}
\usepackage{bm}
\usepackage{caption}
\usepackage{booktabs}
\date{}

\usepackage{booktabs}
\usepackage{longtable}
\usepackage{array}
\usepackage{multirow}
\usepackage{wrapfig}
\usepackage{float}
\usepackage{colortbl}
\usepackage{pdflscape}
\usepackage{tabu}
\usepackage{threeparttable}
\usepackage{threeparttablex}
\usepackage[normalem]{ulem}
\usepackage{makecell}
\usepackage{xcolor}

\begin{document}

\begin{frontmatter}

	\title{Individualized treatment effect was predicted best by modeling baseline risk in interaction with treatment assignment}
    \author[1]{Alexandros Rekkas\corref{cor1}}
  
    \author[1]{Peter R. Rijnbeek}
  
    \author[2]{David M. Kent}
  
    \author[3]{Ewout W. Steyerberg}
  
    \author[4]{David van Klaveren}
  
      \affiliation[1]{Department of Medical Informatics, Erasmus Medical
Center, Rotterdam, The Netherlands}
    \affiliation[2]{Predictive Analytics and Comparative Effectiveness
Center, Institute for Clinical Research and Health Policy Studies, Tufts
Medical Center, Boston, Massachusetts, USA}
    \affiliation[3]{Department of Biomedical Data Sciences, Leiden
University Medical Center, Leiden, The Netherlands}
    \affiliation[4]{Department of Public Health, Erasmus Medical Center,
Rotterdam, The Netherlands}
    \cortext[cor1]{Corresponding author}

  \begin{abstract}
  \textbf{Objective}: To compare different risk-based methods for
  optimal prediction of treatment effects. \textbf{Study Design and
  Setting}: We simulated RCT data using diverse assumptions for the
  average treatment effect, a baseline prognostic index of risk (PI),
  the shape of its interaction with treatment (none, linear, quadratic
  or non-monotonic), and the magnitude of treatment-related harms (none
  or constant independent of the PI). We predicted absolute benefit
  using: models with a constant relative treatment effect;
  stratification in quarters of the PI; models including a linear
  interaction of treatment with the PI; models including an interaction
  of treatment with a restricted cubic spline (RCS) transformation of
  the PI; an adaptive approach using Akaike's Information Criterion. We
  evaluated predictive performance using root mean squared error and
  measures of discrimination and calibration for benefit.
  \textbf{Results}: The linear-interaction model displayed optimal or
  close-to-optimal performance across many simulation scenarios with
  moderate sample size (N=4,250; \textasciitilde785 events). The
  RCS-model was optimal for strong non-linear deviations from a constant
  treatment effect, particularly when sample size was larger (N=17,000).
  The adaptive approach also required larger sample sizes. These
  findings were illustrated in the GUSTO-I trial. \textbf{Conclusion}:
  An interaction between baseline risk and treatment assignment should
  be considered to improve treatment effect predictions.
  \end{abstract}
    \begin{keyword}
    treatment effect heterogeneity \sep absolute benefit \sep 
    prediction models
  \end{keyword}
  
 \end{frontmatter}

\hypertarget{introduction}{%
\section{Introduction}\label{introduction}}

Predictive approaches to heterogeneity of treatment effects (HTE) aim at
the development of models predicting either individualized effects or
which of two (or more) treatments is better for an individual
\citep{Varadhan2013}. In prior work, we divided such methods in three
broader categories based on the reference class used for defining
patient similarity when making individualized predictions or
recommendations \citep{Rekkas2020}. Risk-modeling approaches use
prediction of baseline risk as the reference; treatment effect modeling
approaches also model treatment-covariate interactions, in addition to
risk factors; optimal treatment regime approaches focus on developing
treatment assignment rules and rely heavily on modeling treatment effect
modifiers.

Risk-modeling approaches to predictive HTE analyses provide a viable
option in the absence of well-established treatment effect modifiers
\citep{Kent2019, PathEnE}. In simulations, modeling treatment-covariate
interactions, often led to miscalibrated predictions of absolute
benefit, contrary to risk-based methods, despite their weaker
discrimination of benefit in the presence of true effect modifiers
\citep{vanKlaveren2019}. Most often, risk-modeling approaches are
carried out in two steps: first a risk prediction model is developed
externally or internally on the entire RCT population, ``blinded'' to
treatment; then the RCT population is stratified using this prediction
model to evaluate risk-based treatment effect variation
\citep{Kent2010}. This approach identified substantial absolute
treatment effect differences between low-risk and high-risk patients in
a re-analysis of 32 large trials \citep{Kent2016}. However, even though
estimates at the risk subgroup level may be accurate, these estimates
may not apply to individual patients.

In the current simulation study, we aim to summarize and compare
different risk-based models for predicting treatment effects. We
simulate different relations between baseline risk and treatment effects
and also consider potential harms of treatment. We illustrate the
different models by a case study of predicting individualized effects of
treatment for acute myocardial infarction (MI) in a large RCT.

\hypertarget{methods}{%
\section{Methods}\label{methods}}

\hypertarget{notation}{%
\subsection{Notation}\label{notation}}

We observe RCT data \((Z, X, Y)\), where for each patient \(Z_i= 0, 1\)
is the treatment status, \(Y_i = 0, 1\) is the observed outcome and
\(X_i\) is a set of measured covariates. Let \(\{Y_i(z), z=0, 1\}\)
denote the unobservable potential outcomes. We observe
\(Y_i = Z_iY_i(1) + (1 - Z_i)Y_i(0)\). We are interested in predicting
the conditional average treatment effect (CATE),
\[\tau(x) = E\{Y(0) - Y(1)|X=x\}\] Assuming that
\(\big(Y(0), Y(1)\big)\perp \!\!\! \perp Z|X\), as we are in the RCT
setting, we can predict CATE from \begin{align*}
\tau(x) &= E\{Y(0)\:\vert\:X=x\}-E\{Y(1)\:\vert\:X=x\}\\
&=E\{Y\:\vert\:X=x, Z=0\}-E\{Y\:\vert\:X=x, Z=1\}
\end{align*}

\hypertarget{simulation-scenarios}{%
\subsection{Simulation scenarios}\label{simulation-scenarios}}

We simulated a typical RCT, comparing equally-sized treatment and
control arms in terms of a binary outcome. For each patient we generated
8 baseline covariates \(x_1,\dots,x_4\sim N(0, 1)\) and
\(x_5,\dots,x_8\sim B(1,0.2)\). Outcomes in the control arm were
generated from Bernoulli variables with true probabilities following a
logistic regression model including all baseline covariates, i.e.
\(P(Y(0)=1\:\vert\:X=x) = \text{expit}(lp_0) = e^{lp_0}/(1+e^{lp_0})\),
with \(lp_0=lp_0(x)=x^t\beta\). In the base scenarios coefficient values
\(\beta\) were such, that the AUC of the logistic regression model was
0.75 and the event rate in the control arm was \(20\%\).

Outcomes in the treatment arm were first generated using 3 simple
scenarios: absent (OR = 1), moderate (OR = 0.8) or strong (OR = 0.5)
constant relative treatment effect. We then introduced linear, quadratic
and non-monotonic deviations from constant treatment effects using:
\[lp_1 = \gamma_2(lp_0-c)^2 + \gamma_1(lp_0-c) + \gamma_0, \] where
\(lp_1\) is the true linear predictor in the treatment arm, so that
\(P(Y(1)=1\:\vert\:X=x) = \text{expit}(lp_1)\). Finally, we incorporated
constant absolute harms for all treated patients, such that
\(P(Y(1)=1|X=x) = \text{expit}(lp_1) + \text{harm}\).

The sample size for the base scenarios was set to 4,250 (80\% power for
the detection of a marginal OR of 0.8 with the standard alpha of 5\%).
We evaluated the effect of smaller or larger sample sizes of 1,063 and
17,000, respectively. We also evaluated the effect of risk model
discriminative ability, adjusting the baseline covariate coefficients,
such that the AUC of the regression model in the control arm was 0.65
and 0.85, respectively.

These settings resulted in a simulation study of 648 scenarios covering
the HTE observed in 32 large trials as well as many other potential
variations of risk-based treatment effect (Supplement, Sections 2 and 3)
\citep{Kent2016}.

\hypertarget{individualized-risk-based-benefit-predictions}{%
\subsection{Individualized risk-based benefit
predictions}\label{individualized-risk-based-benefit-predictions}}

In each simulation run we internally developed a prediction model on the
entire population, using a logistic regression with main effects for all
baseline covariates and treatment assignment. Individual risk
predictions were derived by setting treatment assignment to 0. Another
approach would be to derive the prediction model solely on the control
patients; however, it has been shown to lead to biased benefit
predictions \citep{vanKlaveren2019, Burke2014, Abadie2018}.

A \emph{stratified HTE method} has been suggested as an alternative to
traditional subgroup analyses \citep{Kent2019, PathEnE}. Patients are
stratified into equally-sized risk strata---in this case based on risk
quartiles. Absolute treatment effects within risk strata are estimated
by the difference in event rate between control and treatment arm
patients. We considered this approach as a reference, expecting it to
perform worse than the other candidates, as its objective is to provide
an illustration of HTE rather than to optimize individualized benefit
predictions.

Second, we considered a model which assumes \emph{constant relative
treatment effect} (constant odds ratio). Hence, absolute benefit is
predicted from
\(\tau(x;\hat{\beta}) = \text{expit}(\hat{lp}_0) - \text{expit}(\hat{lp}_0+\delta_0)\),
where \(\delta_0\) is the log of the assumed constant odds ratio and
\(\hat{lp}_0 = \hat{lp}_0(x;\hat{\beta}) = x^t\hat{\beta}\) the linear
predictor of the estimated baseline risk model.

Third, we considered a logistic regression model including treatment,
the prognostic index, and their linear interaction. Absolute benefit is
then estimated from
\(\tau(x;\hat{\beta}) = \text{expit}(\delta_0+\delta_1\hat{lp}_0) - \text{expit}(\delta_0+\delta_2+(\delta_1+\delta_3)\hat{lp}_0)\)
We will refer to this method as the \emph{linear interaction} approach.

Fourth, we used \emph{restricted cubic splines} (RCS) to relax the
linearity assumption on the effect of the linear predictor
\citep{Harrell1988}. We considered splines with 3 (RCS-3), 4 (RCS-4) and
5 (RCS-5) knots to compare models with different levels of flexibility.

Finally, we considered an adaptive approach using Akaike's Information
Criterion (AIC) for model selection. More specifically, we ranked the
constant relative treatment effect model, the linear interaction model,
and the RCS models with 3, 4, and 5 knots based on their AIC and
selected the one with the lowest value. The extra degrees of freedom
were 1 (linear interaction), 2, 3 and 4 (RCS models) for these
increasingly complex interactions with the treatment effect.

\hypertarget{evaluation-metrics}{%
\subsection{Evaluation metrics}\label{evaluation-metrics}}

We evaluated the predictive accuracy of the considered methods by the
root mean squared error (RMSE):

\[\text{RMSE}=\sqrt{\frac{1}{n}\sum_{i=1}^n\big(\tau(\bm{x}_i) - \hat{\tau}(\bm{x}_i)\big)^2}\]

We compared the discriminative ability of the methods under study using
c-for-benefit and the integrated calibration index (ICI) for benefit
(Supplement, Section 6) \citep{vanKlaveren2018}.

For each scenario we performed 500 replications, within which all the
considered models were fitted. We simulated a super-population of size
500,000 for each scenario within which we calculated RMSE and
discrimination and calibration for benefit of all the models in each
replication.

\hypertarget{empirical-illustration}{%
\subsection{Empirical illustration}\label{empirical-illustration}}

We demonstrated the different methods using 30,510 patients with acute
myocardial infarction (MI) included in the GUSTO-I trial. 10,348
patients were randomized to tissue plasminogen activator (tPA) treatment
and 20,162 were randomized to streptokinase. The outcome of interest was
30-day mortality (total of 2,128 events), recorded for all patients. In
line with previous analyses \citep{Califf1997, Steyerberg2000}, we
fitted a logistic regression model with 6 baseline covariates, i.e.~age,
Killip class, systolic blood pressure, heart rate, an indicator of
previous MI, and the location of MI, to predict 30-day mortality risk
(Supplement, Section 8).

\hypertarget{results}{%
\section{Results}\label{results}}

\hypertarget{simulations}{%
\subsection{Simulations}\label{simulations}}

\begin{figure}
\includegraphics[width=1\linewidth]{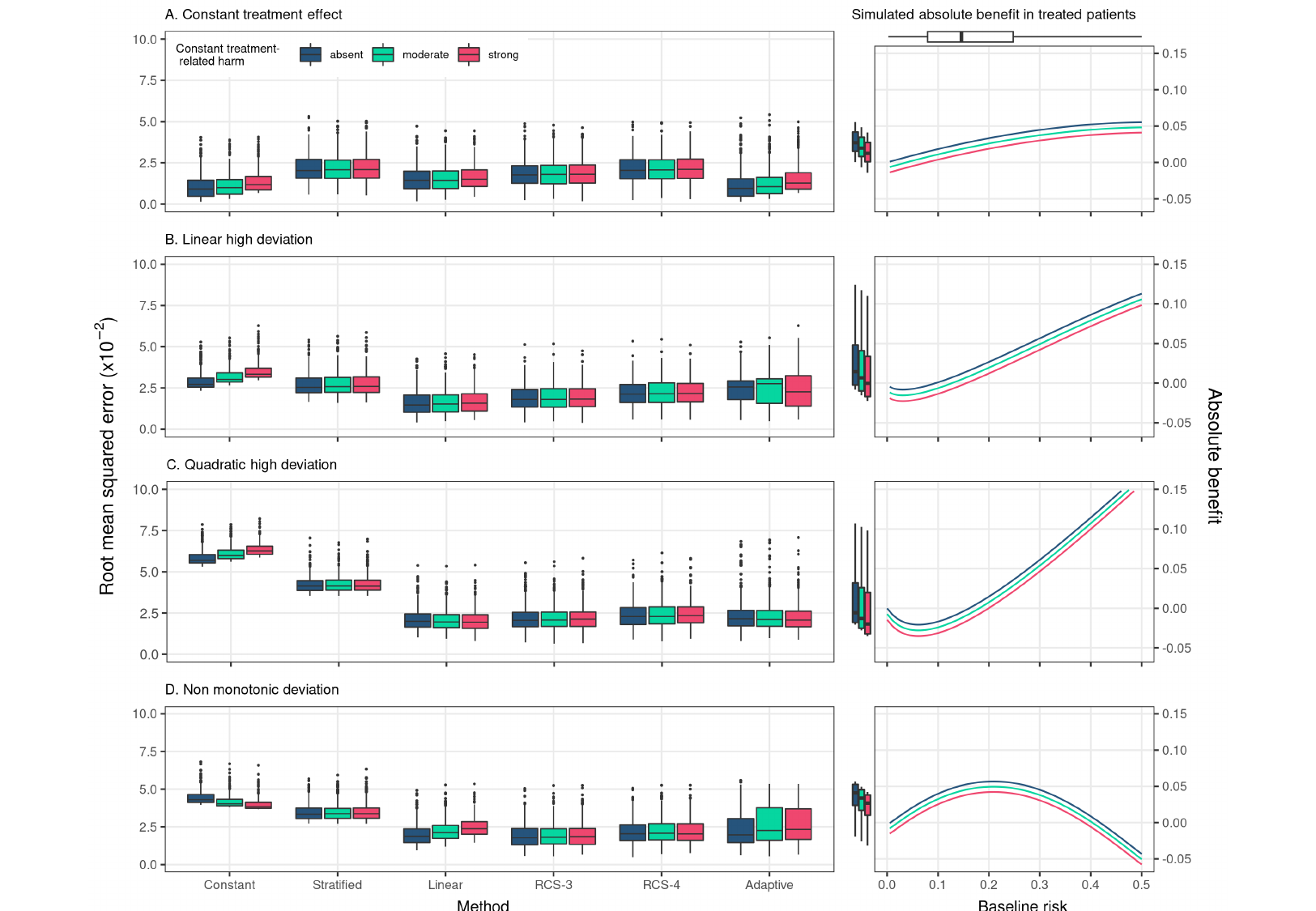} \caption{RMSE of the considered methods across 500 replications calculated from a simulated super-population of size 500,000. The scenario with true constant relative treatment effect (panel A) had a true prediction AUC of 0.75 and sample size of 4250. The RMSE is also presented for strong linear (panel B), strong quadratic (panel C), and non-monotonic (panel D) from constant relative treatment effects. Panels on the right side present the true relations between baseline risk (x-axis) and absolute treatment benefit (y-axis). The 2.5, 25, 50, 75, and 97.5 percentiles of the risk distribution are expressed by the boxplot on the top. The 2.5, 25, 50, 75, and 97.5 percentiles of the true benefit distributions are expressed by the boxplots on the side of the right-handside panel.}\label{fig:rmsebase}
\end{figure}

The constant treatment effect approach outperformed other approaches in
the base case scenario (N = 4,250; OR = 0.8; AUC= 0.75; no absolute
treatment harm) with a true constant treatment effect (median RMSE:
constant treatment effect 0.009; linear interaction 0.014; RCS-3 0.018).
The linear interaction model was optimal under true linear deviations
(median RMSE: constant treatment effect 0.027; linear interaction 0.015;
RCS-3 0.018; Figure \ref{fig:rmsebase} panels A-C) and even in the
presence of true quadratic deviations (median RMSE: constant treatment
effect 0.057; linear interaction 0.020; RCS-3 0.021; Figure
\ref{fig:rmsebase} panels A-C) from a constant relative treatment
effect. With non-monotonic deviations, RCS-3 slightly outperformed the
linear interaction model (median RMSE: linear interaction 0.019; RCS-3
0.018; Figure \ref{fig:rmsebase} panel D). With strong treatment-related
harms the results were very similar in most scenarios (Figure
\ref{fig:rmsebase} panels A-C). Under non-monotonic deviations the
optimal performance of RCS-3 was more pronounced (median RMSE: linear
interaction 0.024; RCS-3 0.019; Figure \ref{fig:rmsebase} panel D). A
stronger average treatment effect (OR=0.5) led to larger absolute
benefit predictions and consequently to larger RMSE for all approaches,
but the relative differences between different approaches were similar
to the base case scenario (Supplement, Figure S10).

The adaptive approach had limited loss of performance in terms of the
median RMSE to the best-performing method in each scenario. However,
compared to the best-performing approach, its RMSE was more variable in
scenarios with linear and non-monotonic deviations, especially when also
including moderate or strong treatment-related harms. On closer
inspection, we found that this behavior was caused by selecting the
constant treatment effect model in a substantial proportion of the
replications (Supplement, Figure S3).

\begin{figure}
\includegraphics[width=1\linewidth]{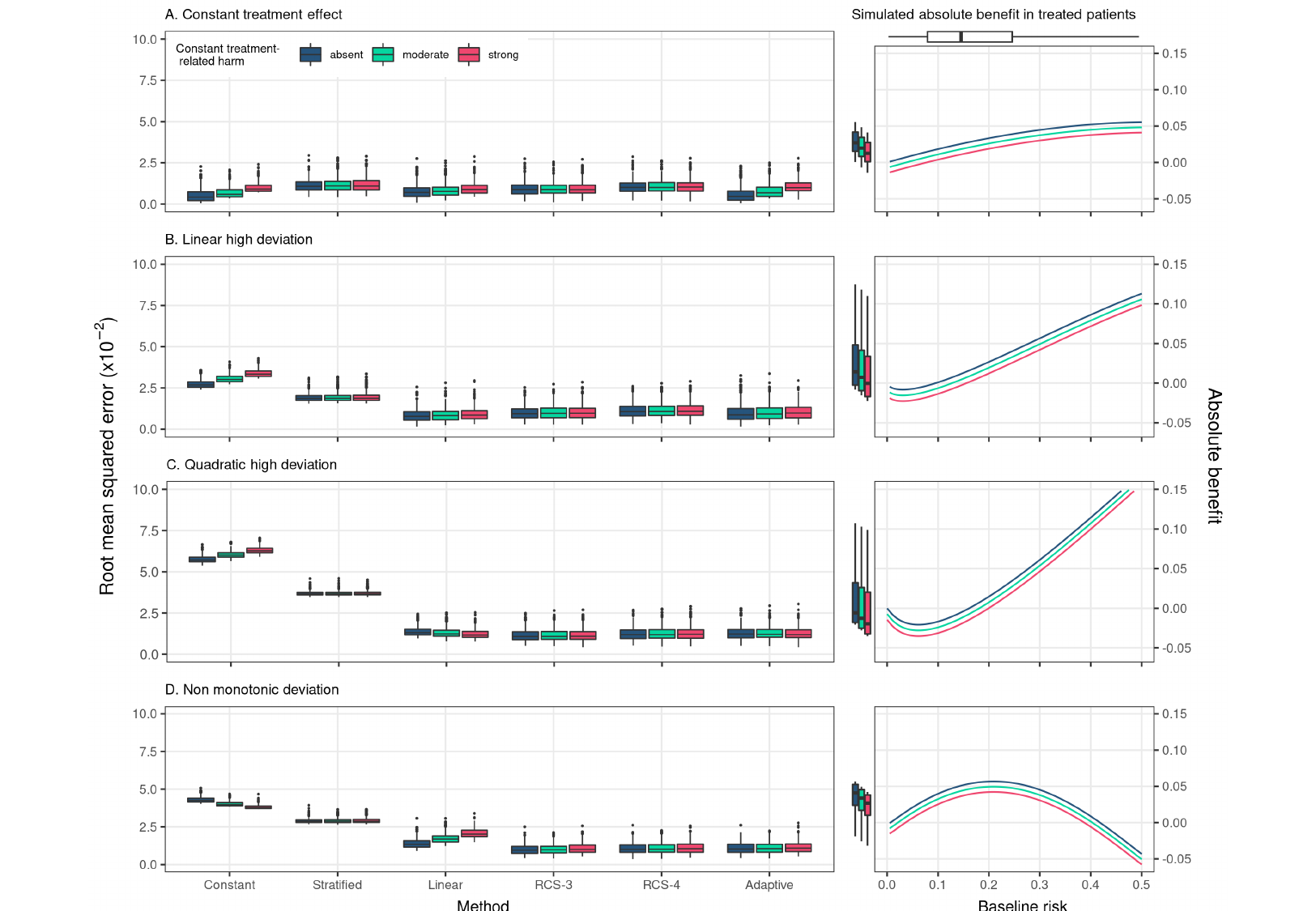} \caption{RMSE of the considered methods across 500 replications calculated in simulated samples of size 17,000 rather than 4,250 in Figure \ref{fig:rmsebase}. RMSE was calculated on a super-population of size 500,000}\label{fig:rmsesamplesize}
\end{figure}

Increasing the sample size to 17,000 favored RCS-3 the most (Figure
\ref{fig:rmsesamplesize}). The difference in performance with the linear
interaction approach was more limited in settings with a constant
treatment effect (median RMSE: linear interaction 0.007; RCS-3 0.009)
and with a true linear interaction (median RMSE: linear interaction
0.008; RCS-3 0.009) and more emphasized in settings with strong
quadratic deviations (median RMSE: linear interaction 0.013; RCS-3
0.011) and non-monotonic deviations (median RMSE: linear interaction
0.014; RCS-3 0.010). Due to the large sample size, the RMSE of the
adaptive approach was even more similar to the best-performing method,
and the constant relative treatment effect model was less often wrongly
selected (Supplement, Figure S4).

\begin{figure}
\includegraphics[width=1\linewidth]{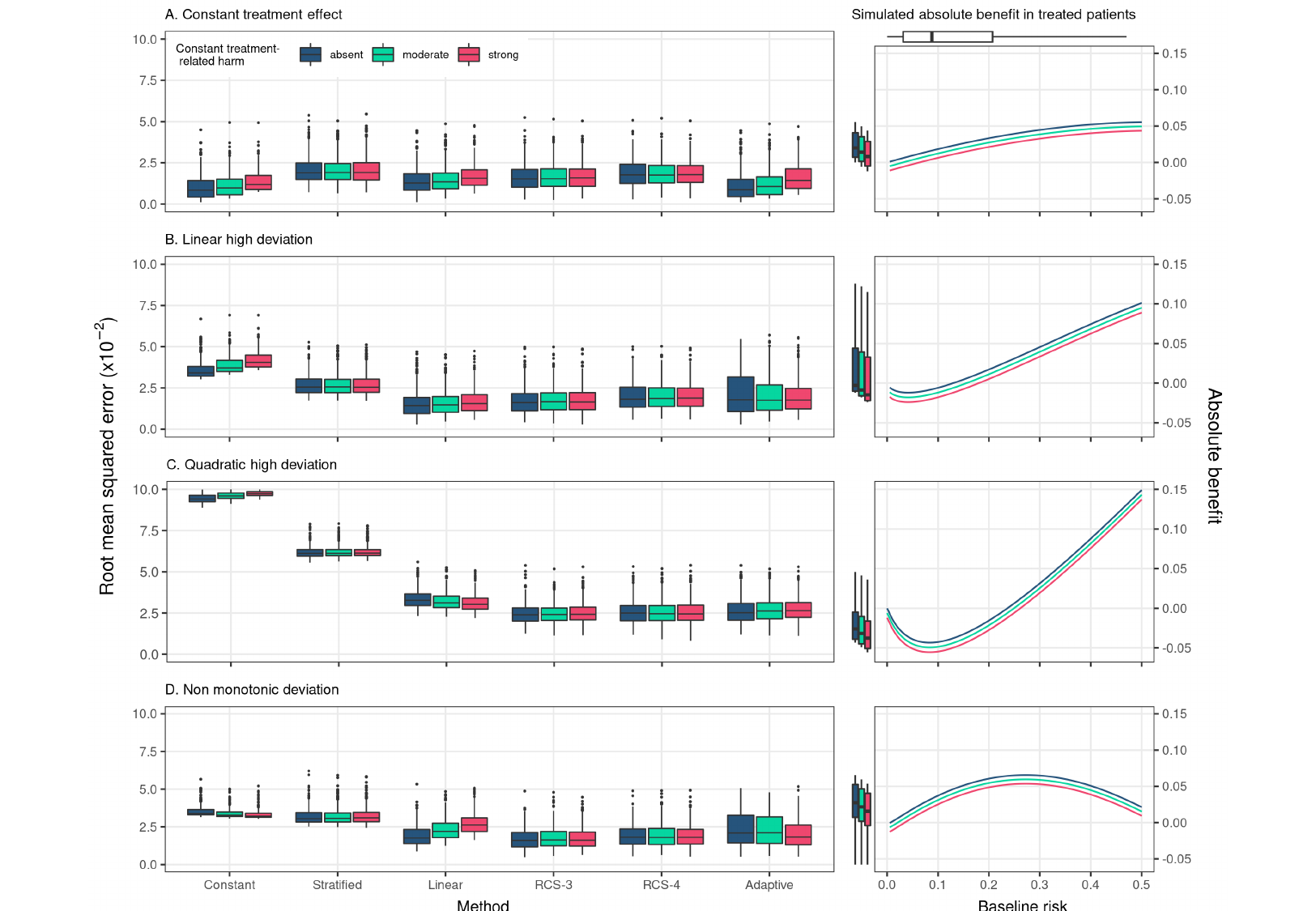} \caption{RMSE of the considered methods across 500 replications calculated in simulated samples 4,250. True prediction AUC of 0.85. RMSE was calculated on a super-population of size 500,000}\label{fig:rmseauc}
\end{figure}

Similarly, when we increased the AUC of the true prediction model to
0.85 (OR = 0.8 and N = 4,250), RCS-3 had the lowest RMSE in the case of
strong quadratic or non-monotonic deviations and very comparable
performance to the -- optimal -- linear interaction model in the case of
strong linear deviations (median RMSE of 0.016 for RCS-3 compared to
0.014 for the linear interaction model; Figure \ref{fig:rmseauc}).
Similar to the base case scenario the adaptive approach wrongly selected
the constant treatment effect model (23\% and 25\% of the replications
in the strong linear and non-monotonic deviation scenarios without
treatment-related harms, respectively), leading to increased variability
of the RMSE (Supplement, Figure S5).

\begin{figure}
\includegraphics[width=1\linewidth]{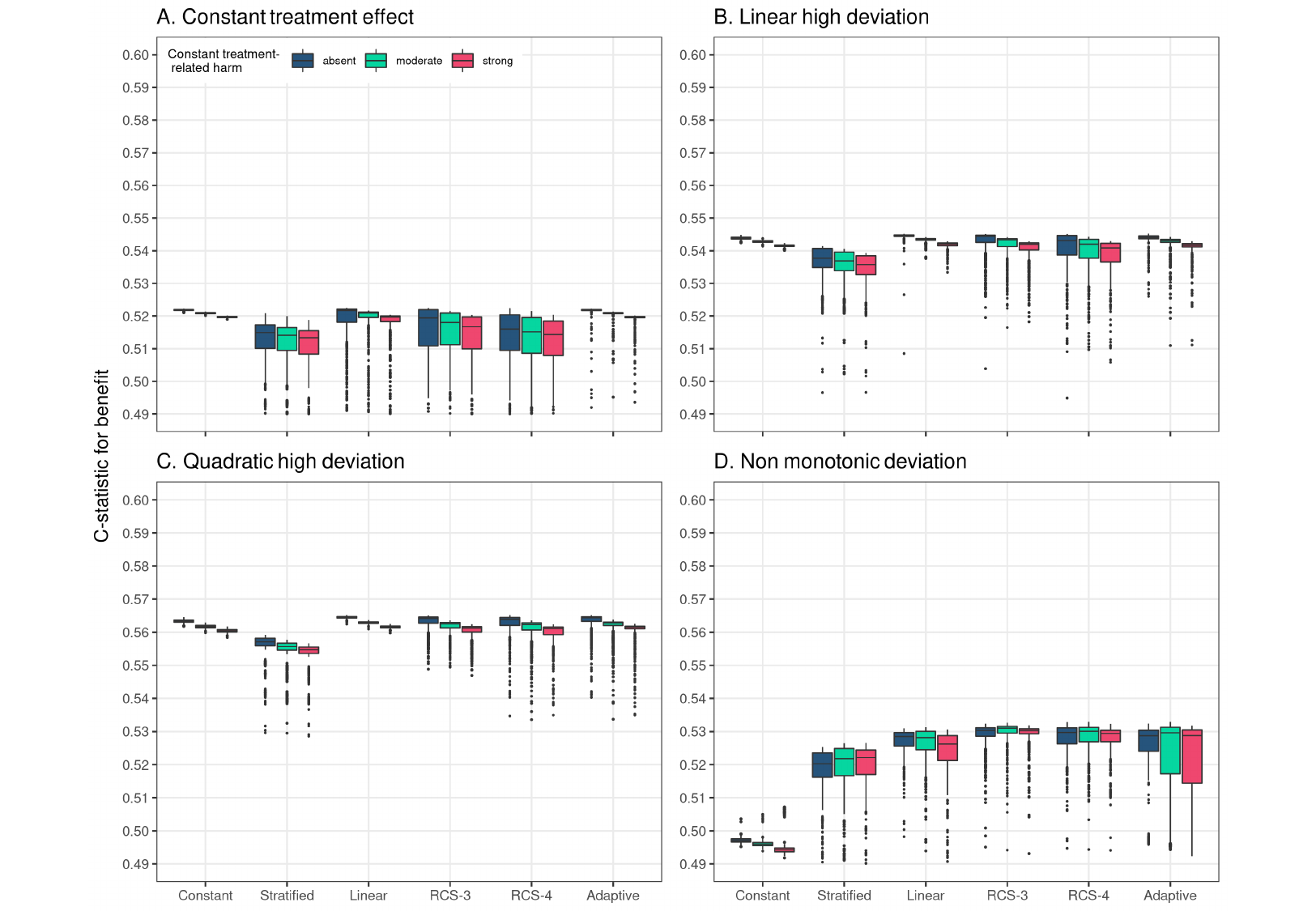} \caption{Discrimination for benefit of the considered methods across 500 replications calculated in a simulated samples of size 4,250. True prediction AUC of 0.75.}\label{fig:discrimination}
\end{figure}

With a true constant relative treatment effect, discrimination for
benefit was only slightly lower for the linear interaction model, but
substantially lower for the non-linear RCS approaches (Figure
\ref{fig:discrimination}; panel A). With strong linear or quadratic
deviations from a constant relative treatment effect, all methods
discriminated quite similarly (Figure \ref{fig:discrimination}; panels
B-C). With non-monotonic deviations, the constant effect model had much
lower discriminative ability compared to all other methods (median AUC
of 0.500 for the constant effects model, 0.528 for the linear
interaction model and 0.530 Figure \ref{fig:discrimination}; panel D).
The adaptive approach was unstable in terms of discrimination for
benefit, especially with treatment-related harms. With increasing number
of RCS knots, we observed decreasing median values and increasing
variability of the c-for-benefit in all scenarios. When we increased the
sample size to 17,000 we observed similar trends, however the
performance of all methods was more stable (Supplement, Figure S6).
Finally, when we increased the true prediction AUC to 0.85 the adaptive
approach was, again, more conservative, especially with non-monotonic
deviations and null or moderate treatment-related harms (Supplement,
Figure S5).

\begin{figure}
\includegraphics[width=1\linewidth]{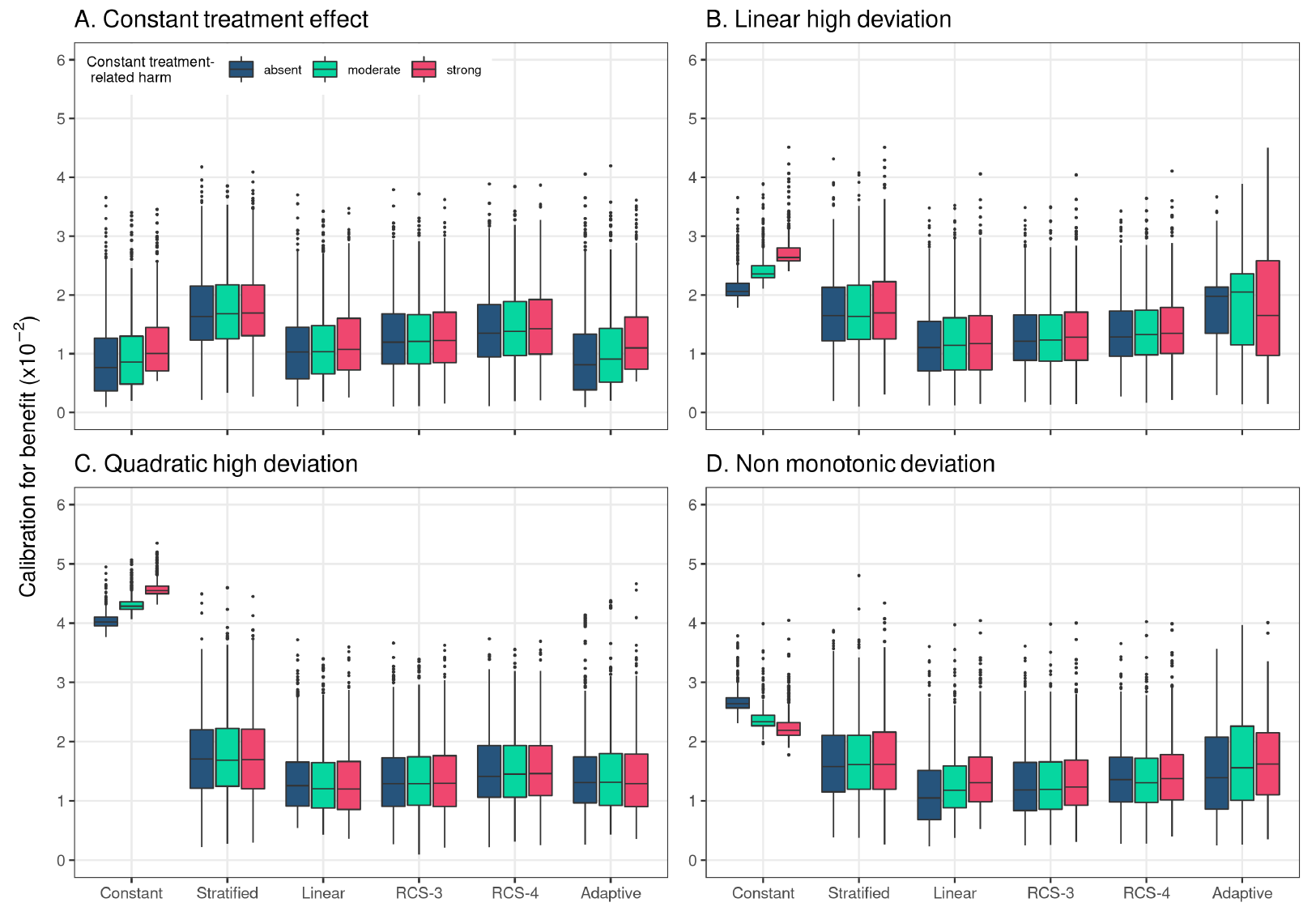} \caption{Calibration for benefit of the considered methods across 500 replications calculated in a simulated sample of size 500,000. True prediction AUC of 0.75 and sample size of 4,250.}\label{fig:calibration}
\end{figure}

In terms of calibration for benefit, the constant effects model
outperformed all other models in the scenario with true constant
treatment effects, but was miscalibrated for all deviation scenarios
(Figure \ref{fig:calibration}). The linear interaction model showed best
or close to best calibration across all scenarios and was only
outperformed by RCS-3 in the case of non-monotonic deviations and
treatment-related harms (Figure \ref{fig:calibration}; panel D). The
adaptive approach was worse calibrated under strong linear and
non-monotonic deviations compared to the linear interaction model and
RCS-3. When we increased the sample size to 17,000 (Supplement, Figure
S6) or the true prediction AUC to 0.85 (Supplement, Figure S7), RCS-3
was somewhat better calibrated than the linear interaction model with
strong quadratic deviations.

The results from all individual scenarios can be explored online at
\url{https://mi-erasmusmc.shinyapps.io/HteSimulationRCT/}. Additionally,
all the code for the simulations can be found at
\url{https://github.com/mi-erasmusmc/HteSimulationRCT}

\hypertarget{empirical-illustration-1}{%
\subsection{Empirical illustration}\label{empirical-illustration-1}}

We used the derived prognostic index to fit a constant treatment effect,
a linear interaction and an RCS-3 model individualizing absolute benefit
predictions. Following our simulation results, RCS-4 and RCS-5 models
were excluded. Finally, an adaptive approach with the 3 candidate models
was applied.

\begin{figure}
\includegraphics[width=1\linewidth]{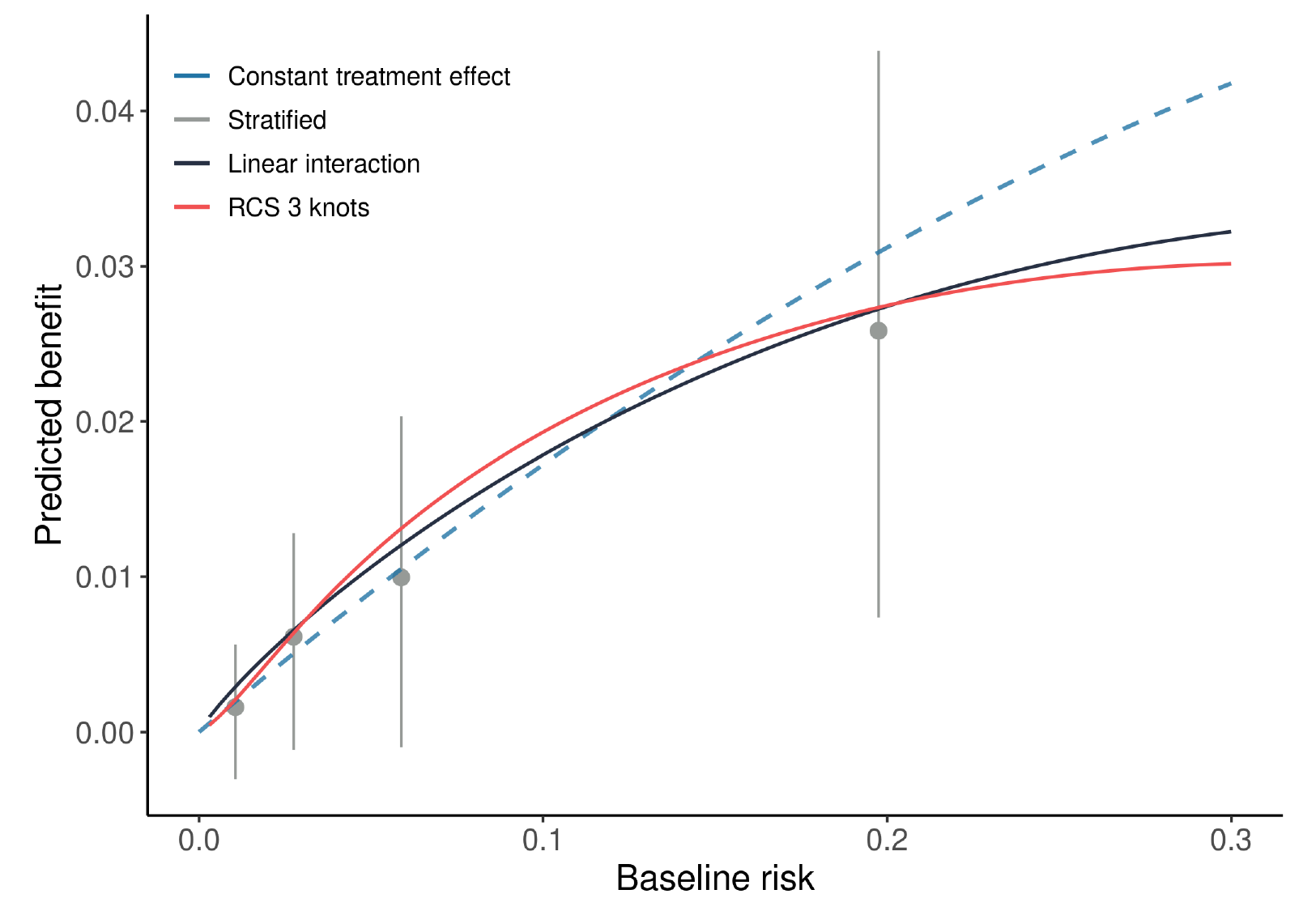} \caption{Individualized absolute benefit predictions based on baseline risk when using a constant treatment effect approach, a linear interaction approach and RCS smoothing using 3 knots. Risk stratified estimates of absolute benefit are presented within quartiles of baseline risk as reference.}\label{fig:gusto}
\end{figure}

All considered methods provided similar fits, predicting increasing
benefits for patients with higher baseline risk predictions, and
followed the evolution of the stratified estimates closely (Figure
\ref{fig:gusto}). The constant treatment effect model had somewhat lower
AIC compared to the linear interaction model (AIC: 9,336 versus 9,342),
equal cross-validated discrimination (c-for-benefit: 0.525), and
slightly better cross-validated calibration (ICI-for benefit: 0.010
versus 0.012). In conclusion, although the sample size (30,510 patients;
2,128 events) allowed for flexible modeling approaches, a simpler
constant treatment effect model is adequate for predicting absolute
30-day mortality benefits of treatment with tPA in patients with acute
MI.

\hypertarget{discussion}{%
\section{Discussion}\label{discussion}}

The linear interaction and the RCS-3 models displayed very good
performance under many of the considered simulation scenarios. The
linear interaction model was optimal in cases with moderate sample sizes
(4.250 patients; \textasciitilde785 events) and moderately performing
baseline risk prediction models, that is, it had lower RMSE, was better
calibrated for benefit and had better discrimination for benefit, even
in scenarios with strong quadratic deviations. In scenarios with true
non-monotonic deviations, the linear interaction model was outperformed
by RCS-3, especially in the presence of treatment-related harms.
Increasing the sample size or the prediction model's discriminative
ability favored RCS-3, especially in scenarios with strong non-linear
deviations from a constant treatment effect.

Our simulation results clearly express the trade-off between the
advantages of flexibly modeling the relationship between baseline risk
and treatment effect and the disadvantages of overfitting this
relationship to the sample at hand. With infinite sample size, the more
flexible approach (here RCS) will be optimal, but in practice, with
limited sample size, parsimonious models may be preferable. Even with
the substantial sample size of our base case scenario, the (less
flexible) linear interaction model performed better than the (more
flexible) RCS approach for most simulation settings. The even less
flexible constant treatment effect model, however, was only optimal when
the treatment effect was truly constant. Moreover, the assumption of a
constant treatment effect may often be too strong
\citep{Rothwell1995, Kent2016}. For example, infants at lower risk of
bronchopulmonary dysplasia benefit relatively more from vitamin A
therapy than infants at higher risk \citep{Rysavy2021}; higher risk
prediabetic patients benefit relatively more from metformin than lower
risk patients \citep{Sussman2015}. Hence, a linear interaction between
baseline risk and the effect of treatment may be the most sensible
approach with moderate sample sizes.

RCS-4 and RCS-5 were too flexible in all considered scenarios, as
indicated by higher RMSE, increased variability of discrimination for
benefit and worse calibration of benefit predictions. Even with larger
sample sizes and strong quadratic or non-monotonic deviations, these
more flexible methods did not outperform the simpler RCS-3 approach.
Higher flexibility may only be helpful under more extreme patterns of
HTE compared to the quadratic deviations considered here. Considering
interactions in RCS-3 models as the most complex approach often may be
reasonable.

Increasing the discriminative ability of the risk model reduced RMSE for
all methods. Higher discrimination translates in higher variability of
predicted risks, which, in turn, allows the considered methods to better
capture absolute treatment benefits. As a consequence, better risk
discrimination also led to higher discrimination between those with low
or high benefit (as reflected in values of c-for-benefit).

The adaptive approach had adequate median performance, following the
``true'' model in most scenarios. With smaller sample sizes it tended to
miss the treatment-baseline risk interaction and selected simpler models
(Supplement Section 4). This conservative behavior resulted in increased
RMSE variability in these scenarios, especially with true strong linear
or non-monotonic deviations. Therefore, with smaller sample sizes the
simpler linear interaction model may be a safer choice for predicting
absolute benefits, especially in the presence of any suspected
treatment-related harms.

One limitation is that we assumed treatment benefit to be a function of
baseline risk in the majority of the simulation scenarios. We attempted
to expand our scenarios by considering moderate and strong constant
treatment-related harms, applied on the absolute scale, in line with
previous work \citep{Glasziou1995}. In a limited set of scenarios with
true interactions between treatment assignment and covariates, our
conclusions remained unchanged (Supplement, Section 7). Even though the
average error rates increased for all the considered methods, due to the
miss-specification of the outcome model, the linear interaction model
had the lowest error rates. RCS-3 had very comparable performance. The
constant treatment effect model was often biased, especially with
moderate or strong treatment-related harms. Future simulation studies
could explore the effect of more extensive deviations from risk-based
treatment effects.

We only focused on risk-based methods, using baseline risk as a
reference in a two-stage approach to individualizing benefit
predictions. However, there is a plethora of different methods, ranging
from treatment effect modeling to tree-based approaches available in
more recent literature
\citep{Athey2019, Lu2018, Wager2018, powers2018some}. Many of these
methods rely on incorporating treatment-covariate interactions when
predicting benefit. An important caveat of such approaches is their
sensitivity to overfitting, which may exaggerate the magnitude of
predicted benefits. In a wide range of simulation settings, a simpler
risk modeling approach was consistently better calibrated for benefit
compared to more complex treatment effect modelling approaches
\citep{vanKlaveren2019}. Similarly, when SYNTAX score II, a model
developed for identifying patients with complex coronary artery disease
that benefit more from percutaneous coronary intervention or from
coronary artery bypass grafting was redeveloped using fewer
treatment-covariate interactions had better external performance
compared to its predecessor
\citep{farooq2013anatomical, takahashi2020redevelopment}. However,
whether this remains the case in a range of empirical settings still
needs to be explored.

In conclusion, the linear interaction approach is a viable option with
moderate sample sizes and/or moderately performing risk prediction
models, assuming a non-constant relative treatment effect plausible.
RCS-3 is a better option with more abundant sample size and when
non-monotonic deviations from a constant relative treatment effect
and/or substantial treatment-related harms are anticipated. Increasing
the complexity of the RCS models by increasing the number of knots does
not improve benefit prediction. Using AIC for model selection is
attractive with larger sample size.

\newpage

\setlength{\parindent}{-0.25in}
\setlength{\leftskip}{0.25in}

\noindent

\hypertarget{refs}{}
\begin{CSLReferences}{0}{0}
\end{CSLReferences}

\setlength{\parindent}{0in}
\setlength{\leftskip}{0in}

\noindent

\bibliography{references.bib}

\begin{thebibliography}{23}
\providecommand{\natexlab}[1]{#1}
\providecommand{\url}[1]{\texttt{#1}}
\expandafter\ifx\csname urlstyle\endcsname\relax
  \providecommand{\doi}[1]{doi: #1}\else
  \providecommand{\doi}{doi: \begingroup \urlstyle{rm}\Url}\fi

\bibitem[Varadhan et~al.(2013)Varadhan, Segal, Boyd, Wu, and
  Weiss]{Varadhan2013}
Ravi Varadhan, Jodi~B. Segal, Cynthia~M. Boyd, Albert~W. Wu, and Carlos~O.
  Weiss.
\newblock A framework for the analysis of heterogeneity of treatment effect
  in~patient-centered outcomes research.
\newblock \emph{Journal of Clinical Epidemiology}, 66\penalty0 (8):\penalty0
  818--825, August 2013.
\newblock \doi{10.1016/j.jclinepi.2013.02.009}.
\newblock URL \url{https://doi.org/10.1016/j.jclinepi.2013.02.009}.

\bibitem[Rekkas et~al.(2020)Rekkas, Paulus, Raman, Wong, Steyerberg, Rijnbeek,
  Kent, and van Klaveren]{Rekkas2020}
Alexandros Rekkas, Jessica~K. Paulus, Gowri Raman, John~B. Wong, Ewout~W.
  Steyerberg, Peter~R. Rijnbeek, David~M. Kent, and David van Klaveren.
\newblock Predictive approaches to heterogeneous treatment effects: a scoping
  review.
\newblock \emph{{BMC} Medical Research Methodology}, 20\penalty0 (1), October
  2020.
\newblock \doi{10.1186/s12874-020-01145-1}.
\newblock URL \url{https://doi.org/10.1186/s12874-020-01145-1}.

\bibitem[Kent et~al.(2019{\natexlab{a}})Kent, Paulus, van Klaveren, D'Agostino,
  Goodman, Hayward, Ioannidis, Patrick-Lake, Morton, Pencina, Raman, Ross,
  Selker, Varadhan, Vickers, Wong, and Steyerberg]{Kent2019}
David~M. Kent, Jessica~K. Paulus, David van Klaveren, Ralph D'Agostino, Steve
  Goodman, Rodney Hayward, John~P.A. Ioannidis, Bray Patrick-Lake, Sally
  Morton, Michael Pencina, Gowri Raman, Joseph~S. Ross, Harry~P. Selker, Ravi
  Varadhan, Andrew Vickers, John~B. Wong, and Ewout~W. Steyerberg.
\newblock The predictive approaches to treatment effect heterogeneity ({PATH})
  statement.
\newblock \emph{Annals of Internal Medicine}, 172\penalty0 (1):\penalty0 35,
  November 2019{\natexlab{a}}.
\newblock \doi{10.7326/m18-3667}.
\newblock URL \url{https://doi.org/10.7326/m18-3667}.

\bibitem[Kent et~al.(2019{\natexlab{b}})Kent, van Klaveren, Paulus, D'Agostino,
  Goodman, Hayward, Ioannidis, Patrick-Lake, Morton, Pencina, Raman, Ross,
  Selker, Varadhan, Vickers, Wong, and Steyerberg]{PathEnE}
David~M. Kent, David van Klaveren, Jessica~K. Paulus, Ralph D'Agostino, Steve
  Goodman, Rodney Hayward, John~P.A. Ioannidis, Bray Patrick-Lake, Sally
  Morton, Michael Pencina, Gowri Raman, Joseph~S. Ross, Harry~P. Selker, Ravi
  Varadhan, Andrew Vickers, John~B. Wong, and Ewout~W. Steyerberg.
\newblock The predictive approaches to treatment effect heterogeneity ({PATH})
  statement: Explanation and elaboration.
\newblock \emph{Annals of Internal Medicine}, 172\penalty0 (1):\penalty0 W1,
  November 2019{\natexlab{b}}.
\newblock \doi{10.7326/m18-3668}.
\newblock URL \url{https://doi.org/10.7326/m18-3668}.

\bibitem[van Klaveren et~al.(2019)van Klaveren, Balan, Steyerberg, and
  Kent]{vanKlaveren2019}
David van Klaveren, Theodor~A. Balan, Ewout~W. Steyerberg, and David~M. Kent.
\newblock Models with interactions overestimated heterogeneity of treatment
  effects and were prone to treatment mistargeting.
\newblock \emph{Journal of Clinical Epidemiology}, 114:\penalty0 72--83,
  October 2019.
\newblock \doi{10.1016/j.jclinepi.2019.05.029}.
\newblock URL \url{https://doi.org/10.1016/j.jclinepi.2019.05.029}.

\bibitem[Kent et~al.(2010)Kent, Rothwell, Ioannidis, Altman, and
  Hayward]{Kent2010}
David~M Kent, Peter~M Rothwell, John~PA Ioannidis, Doug~G Altman, and Rodney~A
  Hayward.
\newblock Assessing and reporting heterogeneity in treatment effects in
  clinical trials: a proposal.
\newblock \emph{Trials}, 11\penalty0 (1), August 2010.
\newblock \doi{10.1186/1745-6215-11-85}.
\newblock URL \url{https://doi.org/10.1186/1745-6215-11-85}.

\bibitem[Kent et~al.(2016)Kent, Nelson, Dahabreh, Rothwell, Altman, and
  Hayward]{Kent2016}
David~M. Kent, Jason Nelson, Issa~J. Dahabreh, Peter~M. Rothwell, Douglas~G.
  Altman, and Rodney~A. Hayward.
\newblock Risk and treatment effect heterogeneity: re-analysis of individual
  participant data from 32 large clinical trials.
\newblock \emph{International Journal of Epidemiology}, page dyw118, July 2016.
\newblock \doi{10.1093/ije/dyw118}.
\newblock URL \url{https://doi.org/10.1093/ije/dyw118}.

\bibitem[Burke et~al.(2014)Burke, Hayward, Nelson, and Kent]{Burke2014}
J.~F. Burke, R.~A. Hayward, J.~P. Nelson, and D.~M. Kent.
\newblock Using internally developed risk models to assess heterogeneity in
  treatment effects in clinical trials.
\newblock \emph{Circulation: Cardiovascular Quality and Outcomes}, 7\penalty0
  (1):\penalty0 163--169, January 2014.
\newblock \doi{10.1161/circoutcomes.113.000497}.
\newblock URL \url{https://doi.org/10.1161/circoutcomes.113.000497}.

\bibitem[Abadie et~al.(2018)Abadie, Chingos, and West]{Abadie2018}
Alberto Abadie, Matthew~M. Chingos, and Martin~R. West.
\newblock Endogenous stratification in randomized experiments.
\newblock \emph{The Review of Economics and Statistics}, 100\penalty0
  (4):\penalty0 567--580, October 2018.
\newblock \doi{10.1162/rest_a_00732}.
\newblock URL \url{https://doi.org/10.1162/rest_a_00732}.

\bibitem[Harrell et~al.(1988)Harrell, Lee, and Pollock]{Harrell1988}
F.~E. Harrell, K.~L. Lee, and B.~G. Pollock.
\newblock Regression models in clinical studies: Determining relationships
  between predictors and response.
\newblock \emph{{JNCI} Journal of the National Cancer Institute}, 80\penalty0
  (15):\penalty0 1198--1202, October 1988.
\newblock \doi{10.1093/jnci/80.15.1198}.
\newblock URL \url{https://doi.org/10.1093/jnci/80.15.1198}.

\bibitem[van Klaveren et~al.(2018)van Klaveren, Steyerberg, Serruys, and
  Kent]{vanKlaveren2018}
David van Klaveren, Ewout~W. Steyerberg, Patrick~W. Serruys, and David~M. Kent.
\newblock The proposed `concordance-statistic for benefit' provided a useful
  metric when modeling heterogeneous treatment effects.
\newblock \emph{Journal of Clinical Epidemiology}, 94:\penalty0 59--68,
  February 2018.
\newblock \doi{10.1016/j.jclinepi.2017.10.021}.
\newblock URL \url{https://doi.org/10.1016/j.jclinepi.2017.10.021}.

\bibitem[Califf et~al.(1997)Califf, Woodlief, Harrell, Lee, White, Guerci,
  Barbash, Simes, Weaver, Simoons, and Topol]{Califf1997}
Robert~M. Califf, Lynn~H. Woodlief, Frank~E. Harrell, Kerry~L. Lee, Harvey~D.
  White, Alan Guerci, Gabriel~I. Barbash, R.John Simes, W.Douglas~D. Weaver,
  Maarten~L. Simoons, and Eric~J. Topol.
\newblock Selection of thrombolytic therapy for individual patients:
  Development of a clinical model.
\newblock \emph{American Heart Journal}, 133\penalty0 (6):\penalty0 630--639,
  June 1997.
\newblock \doi{10.1016/s0002-8703(97)70164-9}.
\newblock URL \url{https://doi.org/10.1016/s0002-8703(97)70164-9}.

\bibitem[Steyerberg et~al.(2000)Steyerberg, Bossuyt, and Lee]{Steyerberg2000}
Ewout~W. Steyerberg, Patrick~M.M. Bossuyt, and Kerry~L. Lee.
\newblock Clinical trials in acute myocardial infarction: Should we adjust for
  baseline characteristics?
\newblock \emph{American Heart Journal}, 139\penalty0 (5):\penalty0 745--751,
  May 2000.
\newblock \doi{10.1016/s0002-8703(00)90001-2}.
\newblock URL \url{https://doi.org/10.1016/s0002-8703(00)90001-2}.

\bibitem[Rothwell(1995)]{Rothwell1995}
P.M. Rothwell.
\newblock Can overall results of clinical trials be applied to all patients?
\newblock \emph{The Lancet}, 345\penalty0 (8965):\penalty0 1616--1619, June
  1995.
\newblock \doi{10.1016/s0140-6736(95)90120-5}.
\newblock URL \url{https://doi.org/10.1016/s0140-6736(95)90120-5}.

\bibitem[Rysavy et~al.(2021)Rysavy, Li, Tyson, Jensen, Das, Ambalavanan,
  Laughon, Greenberg, Patel, Pedroza, and Bell]{Rysavy2021}
Matthew~A. Rysavy, Lei Li, Jon~E. Tyson, Erik~A. Jensen, Abhik Das, Namasivayam
  Ambalavanan, Matthew~M. Laughon, Rachel~G. Greenberg, Ravi~M. Patel, Claudia
  Pedroza, and Edward~F. Bell.
\newblock Should vitamin a injections to prevent bronchopulmonary dysplasia or
  death be reserved for high-risk infants? reanalysis of the national institute
  of child health and human development neonatal research network randomized
  trial.
\newblock \emph{The Journal of Pediatrics}, 236:\penalty0 78--85.e5, September
  2021.
\newblock \doi{10.1016/j.jpeds.2021.05.022}.
\newblock URL \url{https://doi.org/10.1016/j.jpeds.2021.05.022}.

\bibitem[Sussman et~al.(2015)Sussman, Kent, Nelson, and Hayward]{Sussman2015}
J.~B. Sussman, D.~M. Kent, J.~P. Nelson, and R.~A. Hayward.
\newblock Improving diabetes prevention with benefit based tailored treatment:
  risk based reanalysis of diabetes prevention program.
\newblock \emph{{BMJ}}, 350\penalty0 (feb19 2):\penalty0 h454--h454, February
  2015.
\newblock \doi{10.1136/bmj.h454}.
\newblock URL \url{https://doi.org/10.1136/bmj.h454}.

\bibitem[Glasziou and Irwig(1995)]{Glasziou1995}
P.~P Glasziou and L.~M Irwig.
\newblock An evidence based approach to individualising treatment.
\newblock \emph{{BMJ}}, 311\penalty0 (7016):\penalty0 1356--1359, November
  1995.
\newblock \doi{10.1136/bmj.311.7016.1356}.
\newblock URL \url{https://doi.org/10.1136/bmj.311.7016.1356}.

\bibitem[Athey et~al.(2019)Athey, Tibshirani, and Wager]{Athey2019}
Susan Athey, Julie Tibshirani, and Stefan Wager.
\newblock Generalized random forests.
\newblock \emph{The Annals of Statistics}, 47\penalty0 (2), April 2019.
\newblock \doi{10.1214/18-aos1709}.
\newblock URL \url{https://doi.org/10.1214/18-aos1709}.

\bibitem[Lu et~al.(2018)Lu, Sadiq, Feaster, and Ishwaran]{Lu2018}
Min Lu, Saad Sadiq, Daniel~J. Feaster, and Hemant Ishwaran.
\newblock Estimating individual treatment effect in observational data using
  random forest methods.
\newblock \emph{Journal of Computational and Graphical Statistics}, 27\penalty0
  (1):\penalty0 209--219, January 2018.
\newblock \doi{10.1080/10618600.2017.1356325}.
\newblock URL \url{https://doi.org/10.1080/10618600.2017.1356325}.

\bibitem[Wager and Athey(2018)]{Wager2018}
Stefan Wager and Susan Athey.
\newblock Estimation and inference of heterogeneous treatment effects using
  random forests.
\newblock \emph{Journal of the American Statistical Association}, 113\penalty0
  (523):\penalty0 1228--1242, June 2018.
\newblock \doi{10.1080/01621459.2017.1319839}.
\newblock URL \url{https://doi.org/10.1080/01621459.2017.1319839}.

\bibitem[Powers et~al.(2018)Powers, Qian, Jung, Schuler, Shah, Hastie, and
  Tibshirani]{powers2018some}
Scott Powers, Junyang Qian, Kenneth Jung, Alejandro Schuler, Nigam~H Shah,
  Trevor Hastie, and Robert Tibshirani.
\newblock Some methods for heterogeneous treatment effect estimation in high
  dimensions.
\newblock \emph{Statistics in medicine}, 37\penalty0 (11):\penalty0 1767--1787,
  2018.

\bibitem[Farooq et~al.(2013)Farooq, Van~Klaveren, Steyerberg, Meliga, Vergouwe,
  Chieffo, Kappetein, Colombo, Holmes~Jr, Mack, et~al.]{farooq2013anatomical}
Vasim Farooq, David Van~Klaveren, Ewout~W Steyerberg, Emanuele Meliga, Yvonne
  Vergouwe, Alaide Chieffo, Arie~Pieter Kappetein, Antonio Colombo, David~R
  Holmes~Jr, Michael Mack, et~al.
\newblock Anatomical and clinical characteristics to guide decision making
  between coronary artery bypass surgery and percutaneous coronary intervention
  for individual patients: development and validation of syntax score ii.
\newblock \emph{The Lancet}, 381\penalty0 (9867):\penalty0 639--650, 2013.

\bibitem[Takahashi et~al.(2020)Takahashi, Serruys, Fuster, Farkouh, Spertus,
  Cohen, Park, Park, Ahn, Kappetein, et~al.]{takahashi2020redevelopment}
Kuniaki Takahashi, Patrick~W Serruys, Valentin Fuster, Michael~E Farkouh,
  John~A Spertus, David~J Cohen, Seung-Jung Park, Duk-Woo Park, Jung-Min Ahn,
  Arie~Pieter Kappetein, et~al.
\newblock Redevelopment and validation of the syntax score ii to individualise
  decision making between percutaneous and surgical revascularisation in
  patients with complex coronary artery disease: secondary analysis of the
  multicentre randomised controlled syntaxes trial with external cohort
  validation.
\newblock \emph{The Lancet}, 396\penalty0 (10260):\penalty0 1399--1412, 2020.

\end{thebibliography}

\end{document}


\maketitle

{
\setcounter{tocdepth}{2}
\tableofcontents
}
\setcounter{table}{0}  \renewcommand{\thetable}{S\arabic{table}} \setcounter{figure}{0} \renewcommand{\thefigure}{S\arabic{figure}}
\newpage

\hypertarget{notation}{%
\section{Notation}\label{notation}}

We observe RCT data \((Z, X, Y)\), where for each patient \(Z_i= 0, 1\) is the
treatment status, \(Y_i = 0, 1\) is the observed outcome and \(X_i\) is a set of
covariates measured. Let \(\{Y_i(z), z=0, 1\}\) denote the unobservable potential
outcomes. We observe \(Y_i = Z_iY_i(1) + (1 - Z_i)Y_i(0)\). We are interested in
predicting the conditional average treatment effect (CATE),
\[\tau(x) = E\{Y(0) - Y(1)|X=x\}\]
Assuming that \((Z, X, Y)\) is a random sample from the target population and that
\(\big(Y(0), Y(1)\big)\perp \!\!\! \perp Z|X\), as we are in the RCT setting, we can predict
CATE from
\begin{align*}
\tau(x) &= E\{Y(0)\:\vert\:X=x\}-E\{Y(1)\:\vert\:X=x\}\\
&=E\{Y\:\vert\:X=x, Z=0\}-E\{Y\:\vert\:X=x, Z=1\}
\end{align*}

Based on an estimate of baseline risk
\[
E\{Y\:\vert\:X=x, Z=0\}=g\big(\hat{lp}(x)\big)
\]
with \(\hat{u}=\hat{lp}(x)=x^t\hat{\beta}\) the linear predictor and \(g\) the link function,
we predict CATE from
\[
\hat{\tau}(x) = g\big(f(\hat{u}, 0)\big) - g\big(f(\hat{u}, 1)\big)
\]
where \(f(u,z)\) describes interactions of the baseline risk linear predictor with
treatment.

\hypertarget{simulation-settings}{%
\section{Simulation settings}\label{simulation-settings}}

For all patients we observe covariates \(X_1,\dots,X_8\), of which \(4\) are
continuous and \(4\) are binary. More specifically,

\begin{equation*}
X_1,\dots,X_4 \sim N(0, 1)
\end{equation*}
\begin{equation*}
X_5,\dots,X_8 \sim B(1, 0.2)
\end{equation*}

We first, generate the binary outcomes \(Y\) for the untreated patients (\(Z=0\)),
based on

\begin{equation} 
P(Y(0)=1\:\vert\:X=x) = g(\beta_0 + \beta_1x_1+\dots+\beta_8x_8) = g(lp_0),
\label{eq:p0}
\end{equation}

where \[g(x) = \frac{e^x}{1+e^x}\]

For treated patients, outcomes are generated from:

\begin{equation}
P(Y = 1\:\vert\:X=x, Z=1) = g(lp_1)
\label{eq:p1}
\end{equation}

where \[lp_1 = \gamma_2(lp_0-c)^2+\gamma_1(lp_0-c)+\gamma_0\]

\hypertarget{base-case-scenario}{%
\subsection{Base-case scenario}\label{base-case-scenario}}

The base-case scenario assumes a constant odds ratio of \(0.8\) in favor of
treatment. The simulated datasets are of size \(n=4250\), where treatment is
allocated at random using a 50/50 split (80\% power for the detection of an
unadjusted OR of 0.8, assuming an event rate of 20\% in the untreated
arm). Outcome incidence in the untreated population is set at \(20\%\). For the
development of the prediction model we use the model defined in \eqref{eq:p0}
including a constant treatment effect. When doing predictions, \(Z\) is set to
\(0\). The value of the true \(\beta\) is such that the above prediction model
has an AUC of \(0.75\).

The previously defined targets are achieved when \(\beta=(-2.08, 0.49,\dots,0.49)^t\). For the derivations in the treatment arm we use
\(\gamma=(\log(0.8), 1, 0)^t\).

\hypertarget{deviations-from-base-case}{%
\subsection{Deviations from base-case}\label{deviations-from-base-case}}

We deviate from the base-case scenario in two ways. First, we alter the overall
target settings of sample size, overall treatment effect and prediction model
AUC. In a second stage, we consider settings that violate the assumption of a
constant relative treatment effect, using a model-based approach.

For the first part, we consider:

\begin{itemize}
\tightlist
\item
  Sample size:

  \begin{itemize}
  \tightlist
  \item
    \(n=1064\)
  \item
    \(n=17000\)
  \end{itemize}
\item
  Overall treatment effect:

  \begin{itemize}
  \tightlist
  \item
    \(OR=0.5\)
  \item
    \(OR=1\)
  \end{itemize}
\item
  Prediction performance:

  \begin{itemize}
  \tightlist
  \item
    \(AUC=0.65\)
  \item
    \(AUC=0.85\)
  \end{itemize}
\end{itemize}

We set the true risk model coefficients to be
\(\bm{\beta} = \big(-1.63,0.26,\dots,0.26\big)^t\) for \(AUC=0.65\) and \(\bm{\beta} = \big(-2.7,0.82,\dots,0.82\big)^t\)
for \(AUC=0.85\). In both cases, \(\beta_0\) is selected so that an event rate of
\(20\%\) is maintained in the control arm.

For the second part linear, quadratic and non-monotonic deviations from the
assumption of constant relative effect are considered. We also consider
different intensity levels of these deviations. Finally, constant absolute
treatment-related harms are introduced, i.e.~positive
(\(0.25\times\text{true average benefit}\)),
strong positive (\(0.50\times\text{true average benefit}\)) and negative
(\(-0.25\times\text{true average benefit}\); i.e.~constant absolute
treatment-related benefit). In case of true absent treatment effects,
treatment-related harms are set to \(1\%, 2\%\) and \(-1\%\) for positive, strong
positive and negative setting, respectively. The settings for these deviations
are defined in Table \ref{tab:tab1}.

\begingroup\fontsize{7}{9}\selectfont

\begin{landscape}
\begin{longtable}[t]{rrrrrrrrrrrrrrrrrrrr}
\caption{\label{tab:tab1}Scenario settings of the entire simulation study.}\\
\toprule
\multicolumn{5}{c}{Analysis ID} & \multicolumn{9}{c}{Baseline risk} & \multicolumn{4}{c}{True treatment effect} & \multicolumn{2}{c}{Benefit} \\
\cmidrule(l{3pt}r{3pt}){1-5} \cmidrule(l{3pt}r{3pt}){6-14} \cmidrule(l{3pt}r{3pt}){15-18} \cmidrule(l{3pt}r{3pt}){19-20}
Scenario & Base & N & AUC & Treatment-related harm & b0 & b1 & b2 & b3 & b4 & b5 & b6 & b7 & b8 & g0 & g1 & g2 & c & Before harms & After harms\\
\midrule
1 & absent & 4,250 & 0.75 & absent & -2.08 & 0.49 & 0.49 & 0.49 & 0.49 & 0.49 & 0.49 & 0.49 & 0.49 & 0.000 & 1.000 & 0.000 & 0 & 0.000 & 0.000\\
2 & absent & 4,250 & 0.75 & moderate-positive & -2.08 & 0.49 & 0.49 & 0.49 & 0.49 & 0.49 & 0.49 & 0.49 & 0.49 & 0.000 & 1.000 & 0.000 & 0 & 0.000 & -0.010\\
3 & absent & 4,250 & 0.75 & strong-positive & -2.08 & 0.49 & 0.49 & 0.49 & 0.49 & 0.49 & 0.49 & 0.49 & 0.49 & 0.000 & 1.000 & 0.000 & 0 & 0.000 & -0.020\\
4 & absent & 4,250 & 0.75 & negative & -2.08 & 0.49 & 0.49 & 0.49 & 0.49 & 0.49 & 0.49 & 0.49 & 0.49 & 0.000 & 1.000 & 0.000 & 0 & 0.000 & 0.010\\
\addlinespace
5 & absent & 4,250 & 0.65 & absent & -1.63 & 0.26 & 0.26 & 0.26 & 0.26 & 0.26 & 0.26 & 0.26 & 0.26 & 0.000 & 1.000 & 0.000 & 0 & 0.000 & 0.000\\
\bottomrule
\end{longtable}
\end{landscape}
\endgroup{}

\newpage

\hypertarget{plausible-scenario-settings}{%
\section{Plausible scenario settings}\label{plausible-scenario-settings}}

In this section we present specific scenarios from our simulation settings in
which evolution of benefit followed similar patterns to {[}1{]}. In this
case patients were stratified into risk quarters based on their true baseline
risk. Within each risk quarter we constructed boxplots of true benefit.

\begin{figure}
\includegraphics[width=0.5\linewidth]{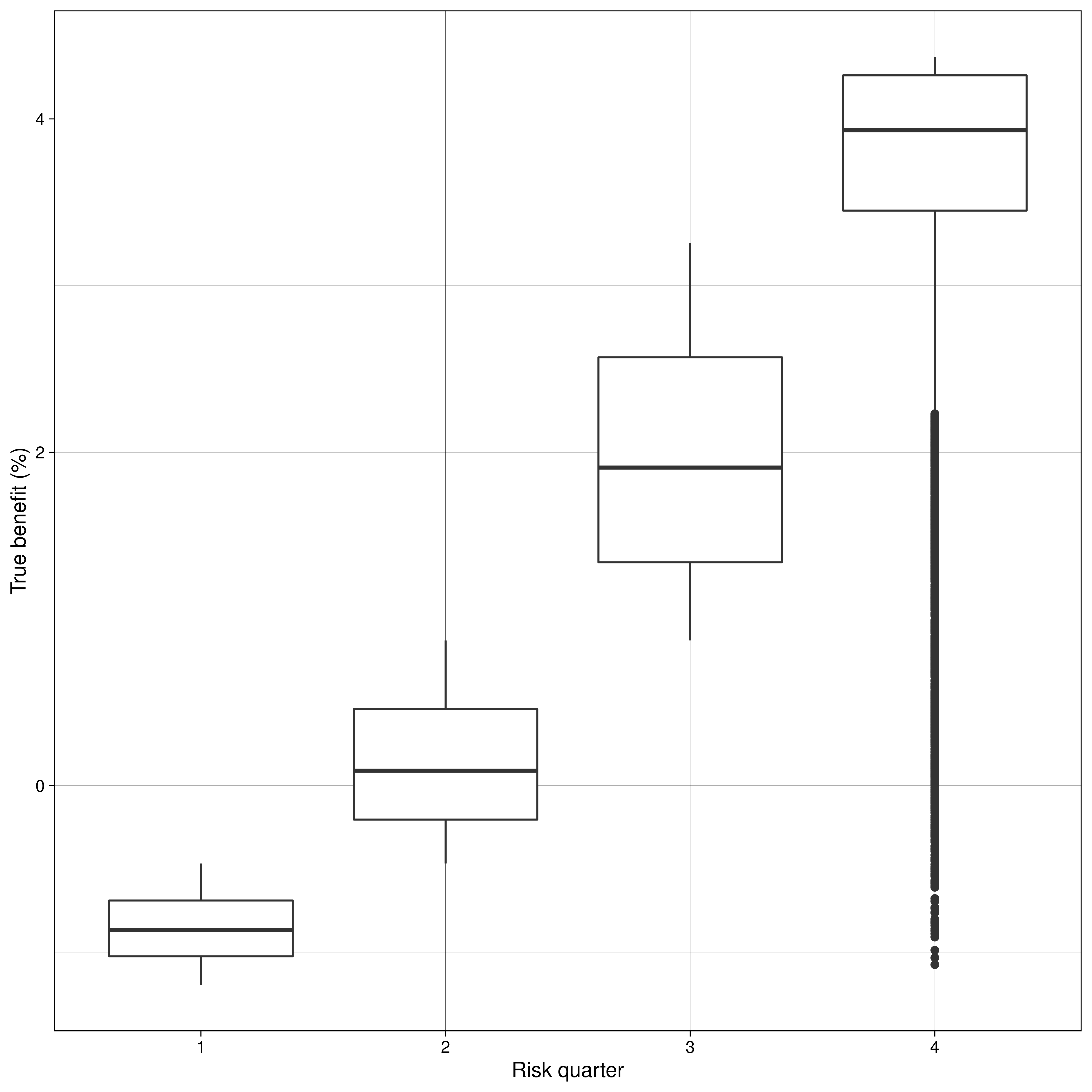} \includegraphics[width=0.5\linewidth]{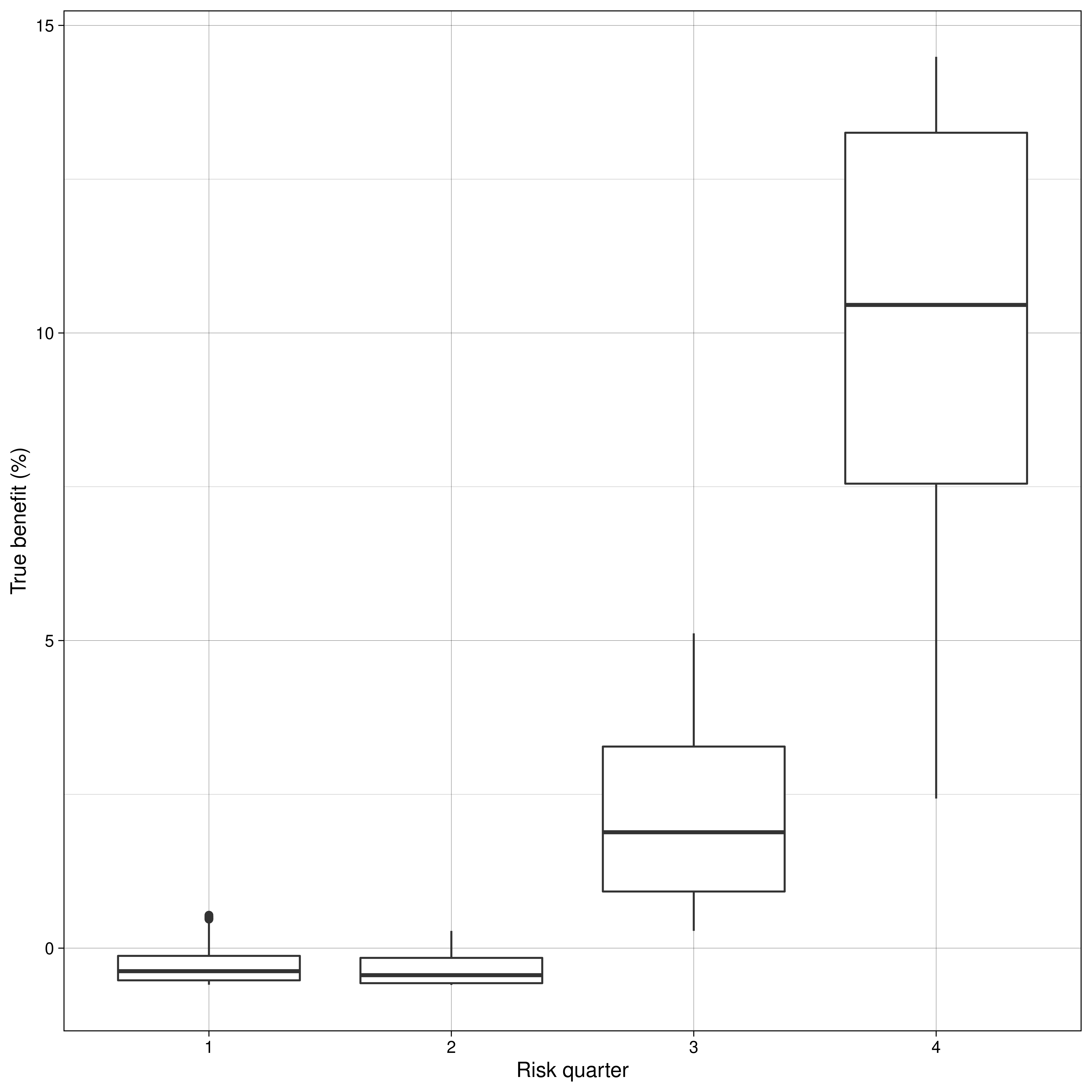} \caption{Simulation scenarios that closely follow trials. In this case, we see increasing absolute benefits with increasing baseline risk.}\label{fig:unnamed-chunk-2}
\end{figure}

\begin{figure}
\includegraphics[width=0.5\linewidth]{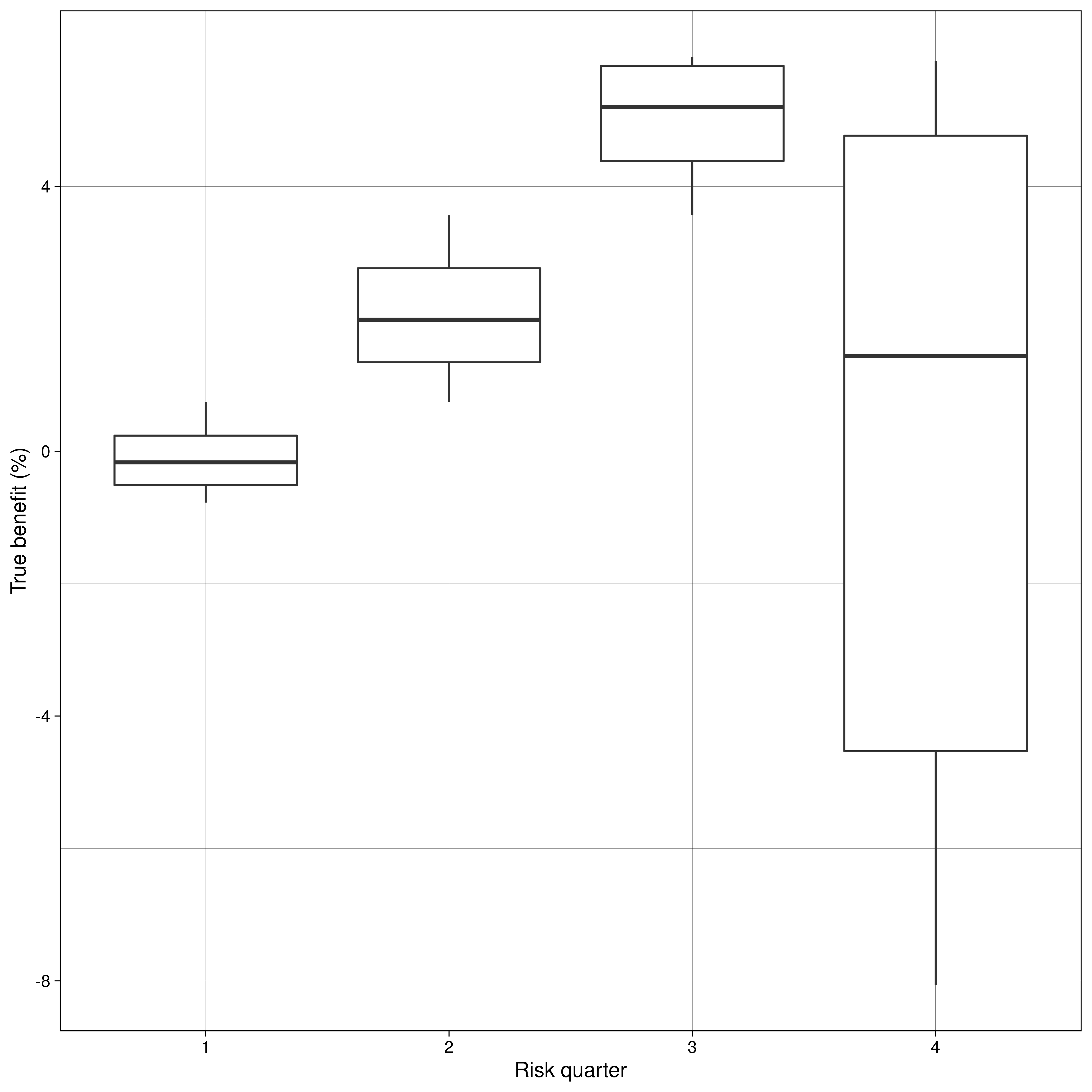} \includegraphics[width=0.5\linewidth]{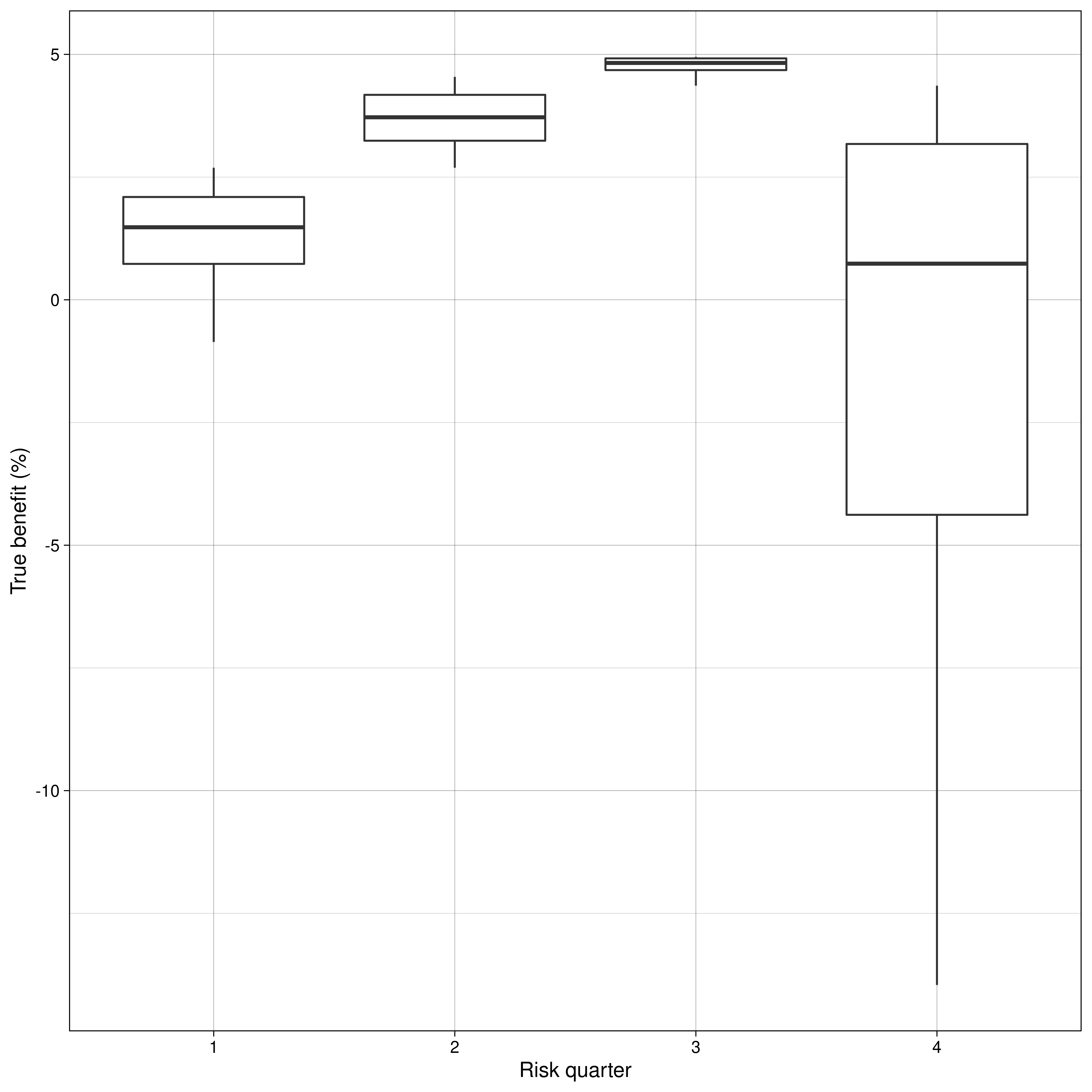} \caption{Simulation scenarios that closely follow trials. In this case, we see increasing absolute benefits with increasing baseline risk up to the third risk quarter. In the fourth risk quarter this trend is interrupted and benefits are diminished.}\label{fig:unnamed-chunk-3}
\end{figure}

\hypertarget{approaches-to-individualize-benefit-predictions}{%
\section{Approaches to individualize benefit predictions}\label{approaches-to-individualize-benefit-predictions}}

\hypertarget{risk-modeling}{%
\subsection{Risk modeling}\label{risk-modeling}}

Merging treatment arms, we develop prediction models including a constant relative treatment effect:

\begin{equation}
P(Y=1\:\vert\:X=x,Z=z) = g(x^t\beta + \delta_0 z)
\end{equation}
\label{eq:risk}

We derive baseline risk predictions for patients by setting \(Z=0\) in
\eqref{eq:risk}. All methods for individualizing benefit predictions are 2-stage
methods, that start by fitting a model for predicting baseline risk. The
estimated linear predictor of this model is

\begin{equation*}
\hat{lp} = lp(x;\hat{\beta}) = x^t\hat{\beta}
\end{equation*}

\hypertarget{risk-stratification}{%
\subsection{Risk stratification}\label{risk-stratification}}

Derive a prediction model using the same approach as above and divide the
population in equally sized risk-based subgroups. Estimate subgroup-specific
absolute benefit from the observed absolute differences. Subject-specific
benefit predictions are made by attributing to individuals their corresponding
subgroup-specific estimate.

\hypertarget{constant-treatment-effect}{%
\subsection{Constant treatment effect}\label{constant-treatment-effect}}

Assuming a constant relative treatment effect, fit the adjusted model in
\eqref{eq:risk}. Then, predict absolute benefit using

\begin{equation}
\hat{\tau}(x;\hat{\beta},\hat{\gamma})=g(f(\hat{lp}, 0)) - g(f(\hat{lp},1)), 
\label{eq:main}
\end{equation}

where \(f(\hat{lp}, z) = \hat{lp}+\hat{\delta}_0z\), with \(\hat{\delta}_0\) the
estimated relative treatment effect (log odds ratio).

\hypertarget{linear-interaction}{%
\subsection{Linear interaction}\label{linear-interaction}}

We relax the assumption of a constant relative treatment effect in
\eqref{eq:main} by setting

\[ f(\hat{lp}, z) = \delta_0+\delta_1z+\delta_2\hat{lp}+\delta_3z\hat{lp} \]

\hypertarget{restricted-cubic-splines}{%
\subsection{Restricted cubic splines}\label{restricted-cubic-splines}}

Finally, we drop the linearity assumption and predict absolute benefit using
smoothing with restricted cubic splines with \(k=3, 4\) and \(5\) knots. More
specifically, we set:

\[ f(\hat{lp}, z) = \delta_0 + \delta_1z+zs(\hat{lp}) \]
where
\[s(x)=\alpha_0+\alpha_1h_1(x)+\alpha_2h_2(x)+\dots+\alpha_{k-1}h_{k-1}(x)\]
with \(h_1(x)=x\) and for \(j=2,\dots,k-2\)
\[h_{j+1}(x)= (x-t_j)^3-(x-t_{k-1})_+^3 \frac{t_k-t_j}{t_k-t_{k-1}}+(x-t_k)^3_+\frac{t_{k-1}-t_j}{t_k-t_{k-1}}\]
where
\(t_1,\dots,t_k\) are the selected knots {[}2{]}.

\newpage

\hypertarget{adaptive-model-selection-frequencies}{%
\section{Adaptive model selection frequencies}\label{adaptive-model-selection-frequencies}}

\begin{figure}
\includegraphics[width=1\linewidth]{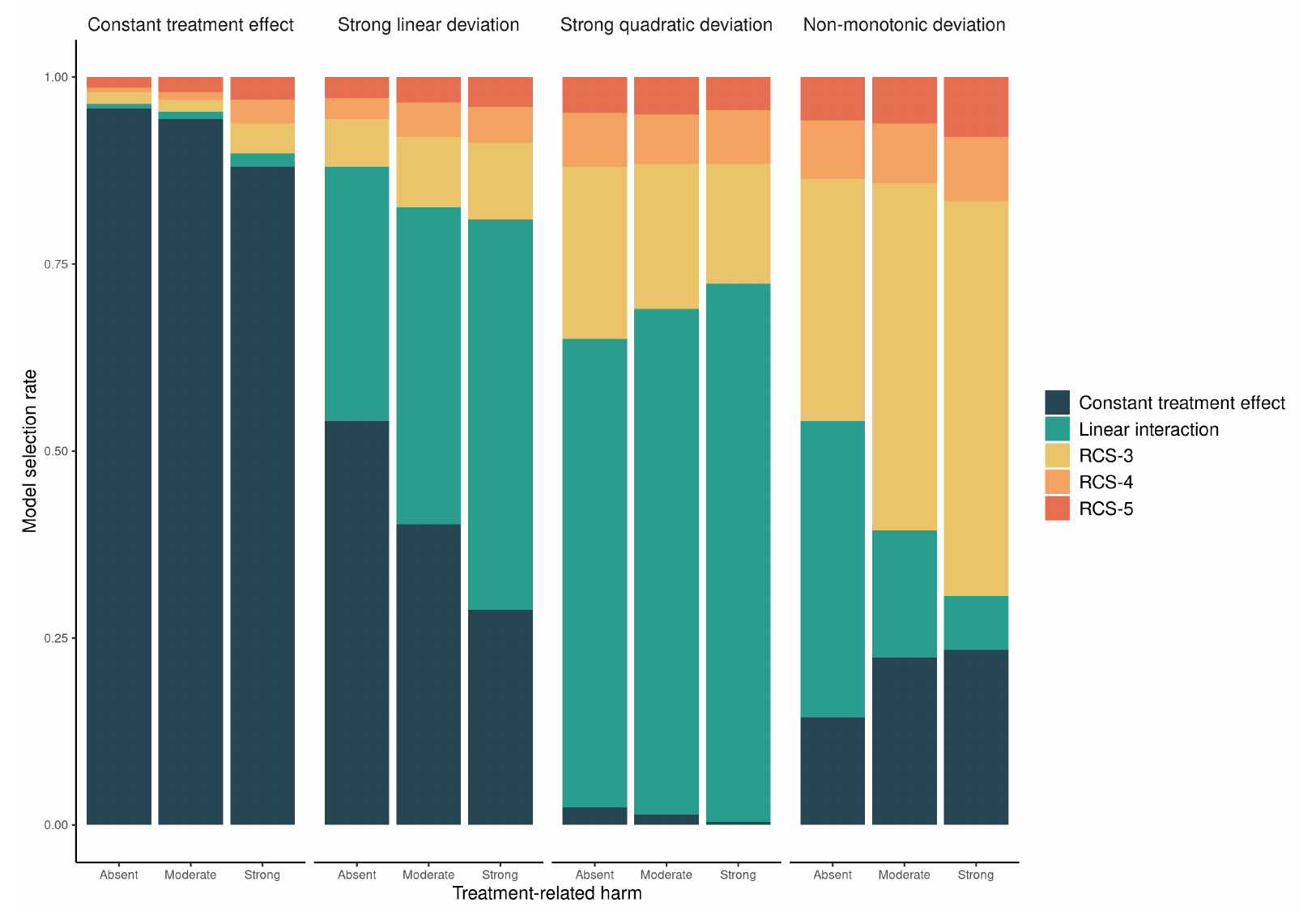} \caption{Model selection frequencies of the adaptive approach based on Akaike's Information Criterion across 500 replications. The scenario with the true constant relative treatment effect (first panel) had a true prediction AUC of 0.75 and sample size of 4,250. }\label{fig:adaptiveBase}
\end{figure}

\begin{figure}
\includegraphics[width=1\linewidth]{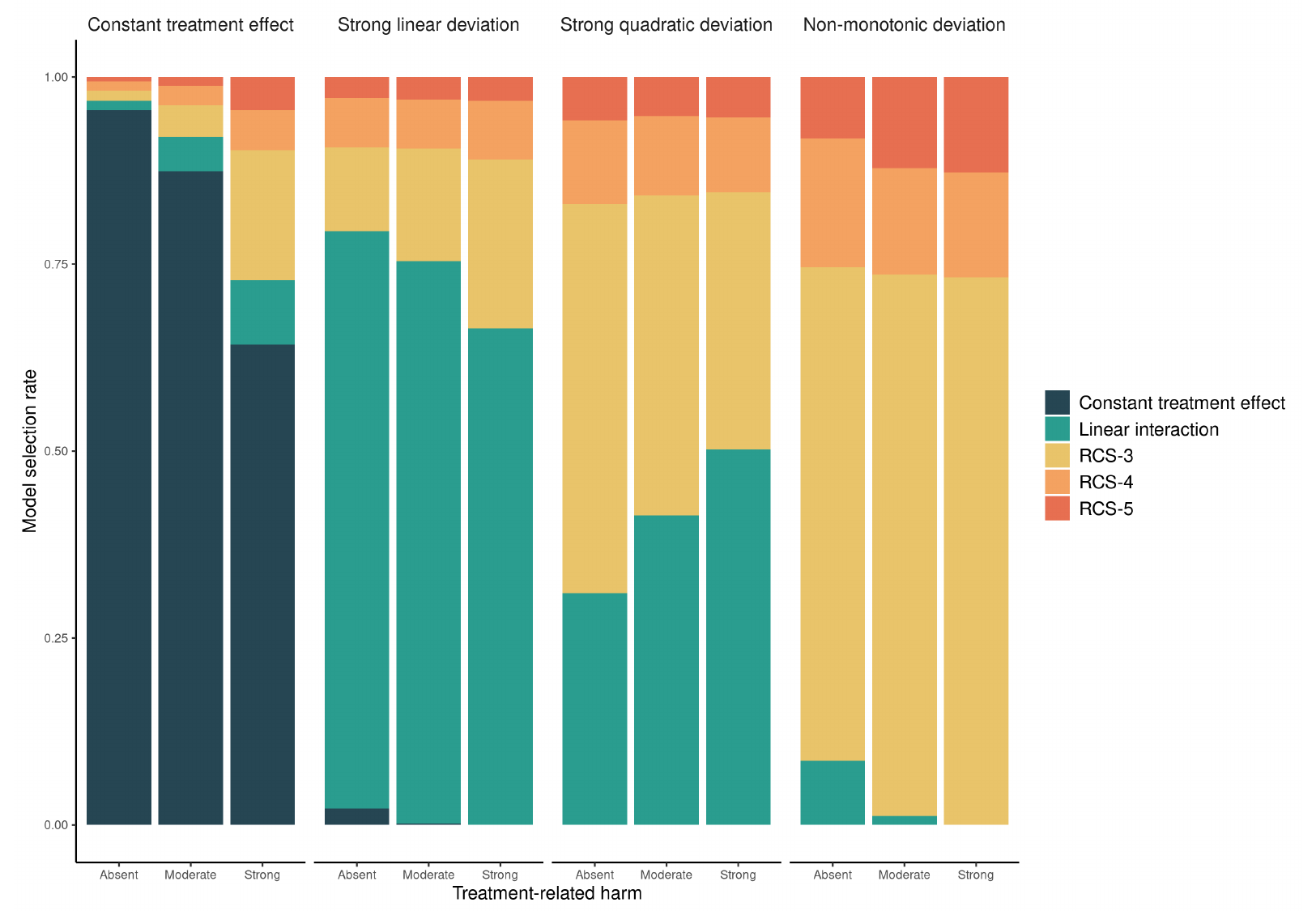} \caption{Model selection frequencies of the adaptive approach based on Akaike's Information Criterion across 500 replications. Sample size is 17,000 rather than 4,250 in Figure \ref{fig:adaptiveBase}}\label{fig:adaptiveSampleSize}
\end{figure}

\begin{figure}
\includegraphics[width=1\linewidth]{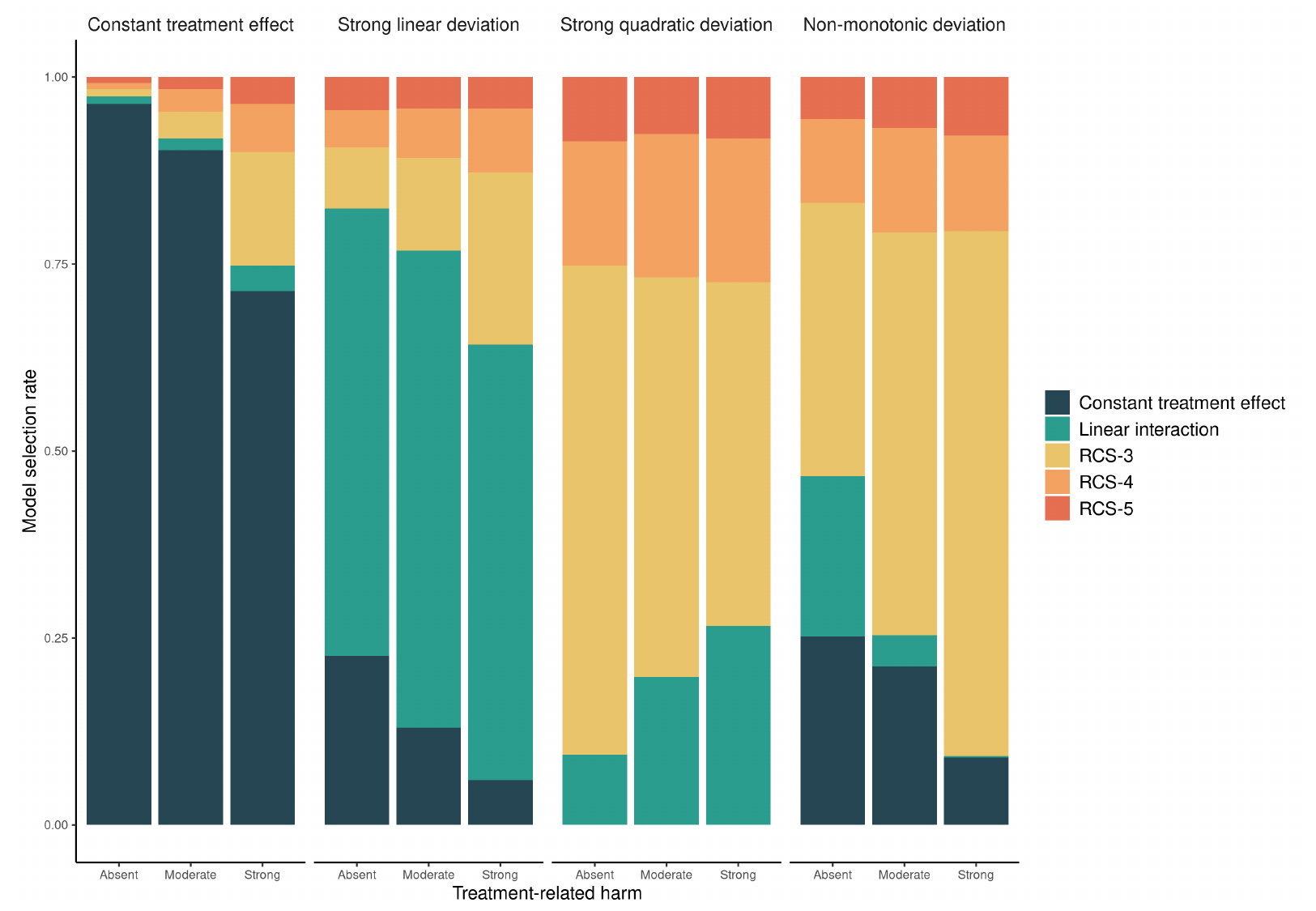} \caption{Model selection frequencies of the adaptive approach based on Akaike's Information Criterion across 500 replications. AUC is 0.85 rather than 0.75 in Figure \ref{fig:adaptiveBase}}\label{fig:adaptiveAuc}
\end{figure}

\newpage

\hypertarget{discrimination-and-calibration-for-benefit}{%
\section{Discrimination and calibration for benefit}\label{discrimination-and-calibration-for-benefit}}

The c-for-benefit represents the probability that from two randomly chosen
matched patient pairs with unequal observed benefit, the pair with greater
observed benefit also has a higher predicted benefit. To be able to calculate
observed benefit, patients in each treatment arm are ranked based on their
predicted benefit and then matched 1:1 across treatment arms. Observed treatment
benefit is defined as the difference of observed outcomes between the untreated
and the treated patient of each matched patient pair. Predicted benefit is
defined as the average of predicted benefit within each matched patient pair.

We evaluated calibration in a similar manner, using the integrated calibration
index (ICI) for benefit {[}3{]}. The observed benefits are regressed on the
predicted benefits using a locally weighted scatterplot smoother (loess). The
ICI-for-benefit is the average absolute difference between predicted and smooth
observed benefit. Values closer to represent better calibration.

\begin{figure}
\includegraphics[width=1\linewidth]{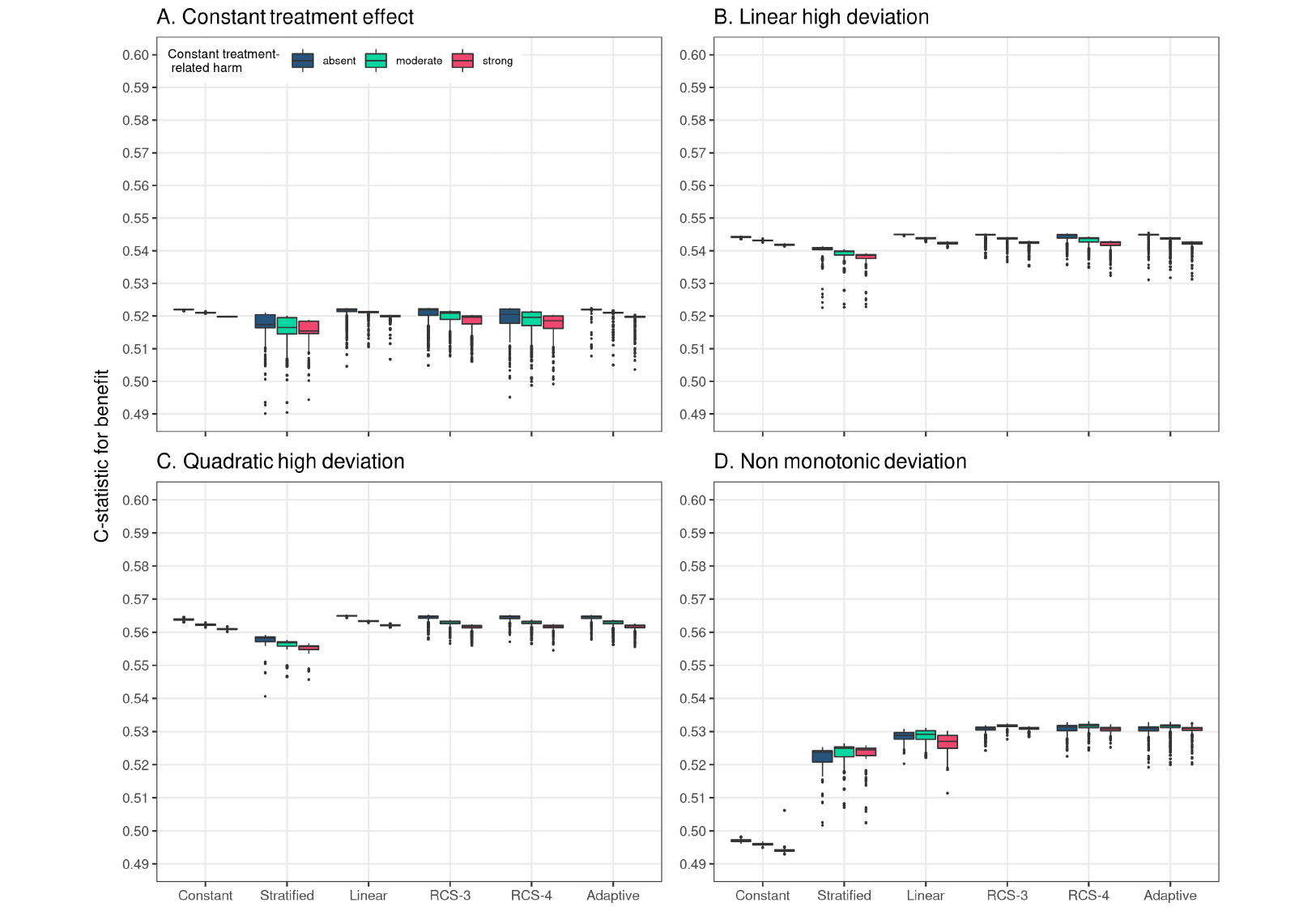} \caption{Discrimination for benefit of the considered methods across 500 replications calculated in a simulated sample of size 500,000. True prediction AUC of 0.75 and sample size of 17,000}\label{fig:discriminationSampleSize}
\end{figure}

\begin{figure}
\includegraphics[width=1\linewidth]{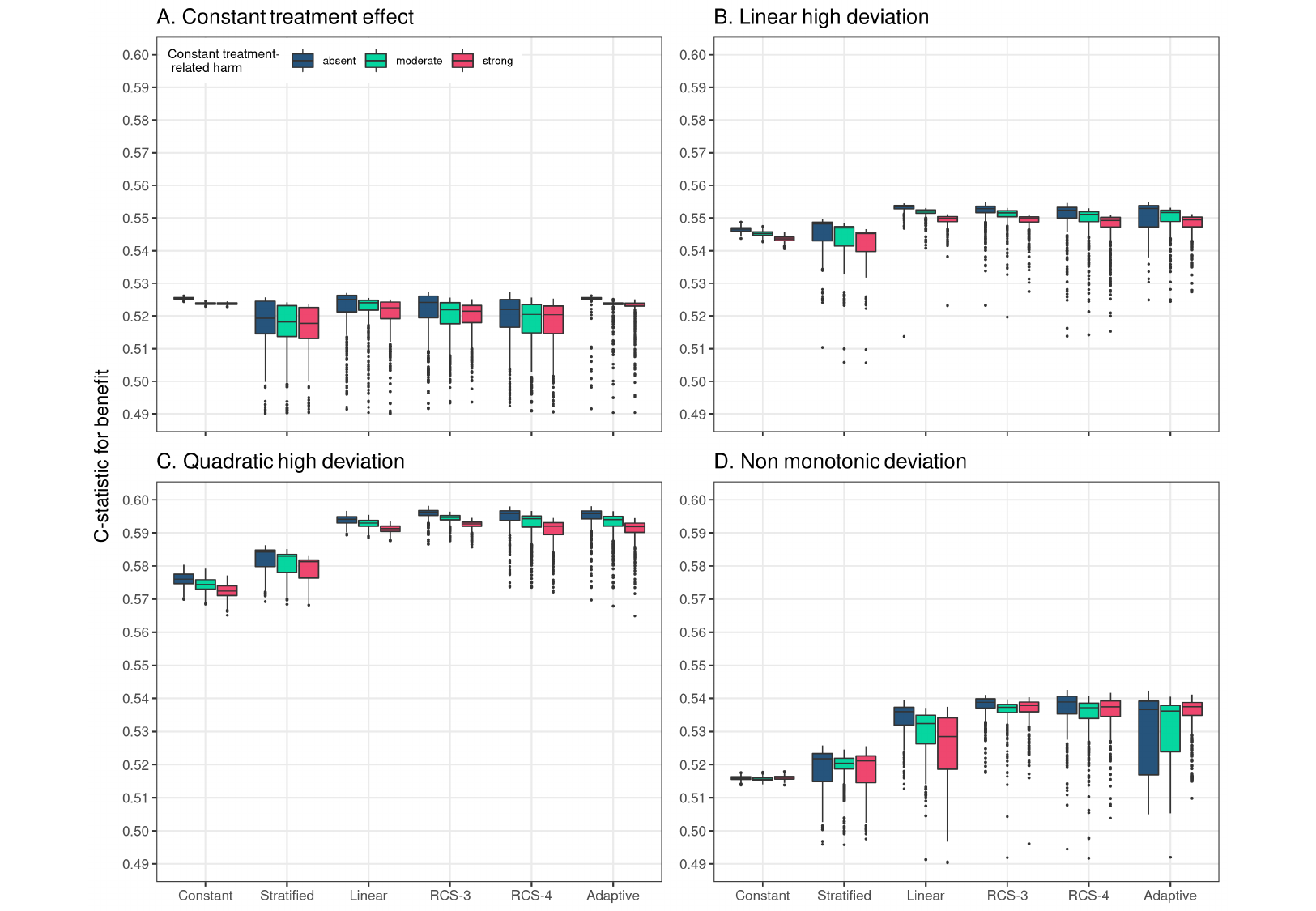} \caption{Discrimination for benefit of the considered methods across 500 replications calculated in a simulated sample of size 500,000. True prediction AUC of 0.85 and sample size of 4,250}\label{fig:discriminationAuc}
\end{figure}

\begin{figure}
\includegraphics[width=1\linewidth]{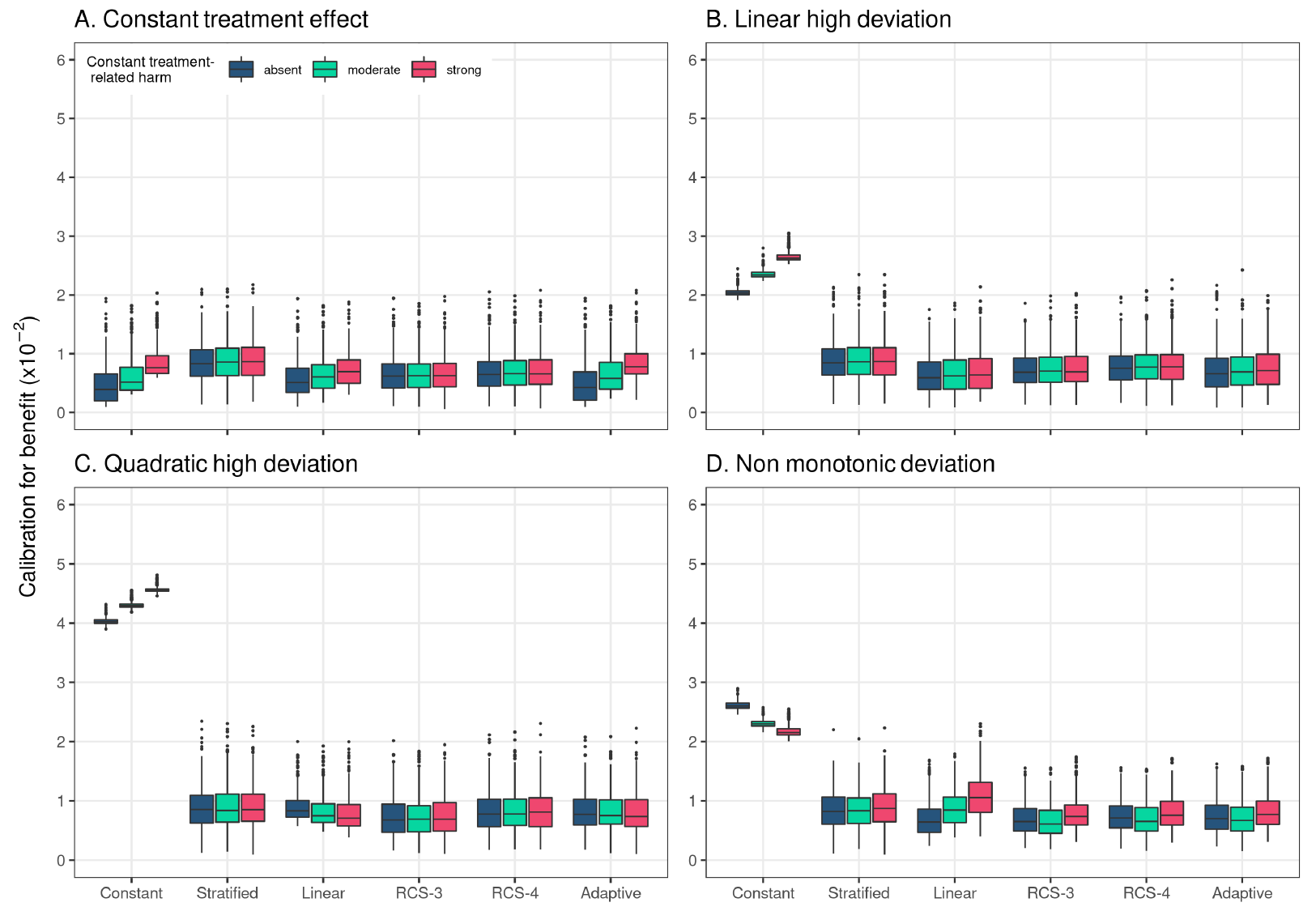} \caption{Calibration for benefit of the considered methods across 500 replications calculated in a simulated sample of size 500,000. True prediction AUC of 0.75 and sample size of 17,000}\label{fig:calibrationSampleSize}
\end{figure}

\begin{figure}
\includegraphics[width=1\linewidth]{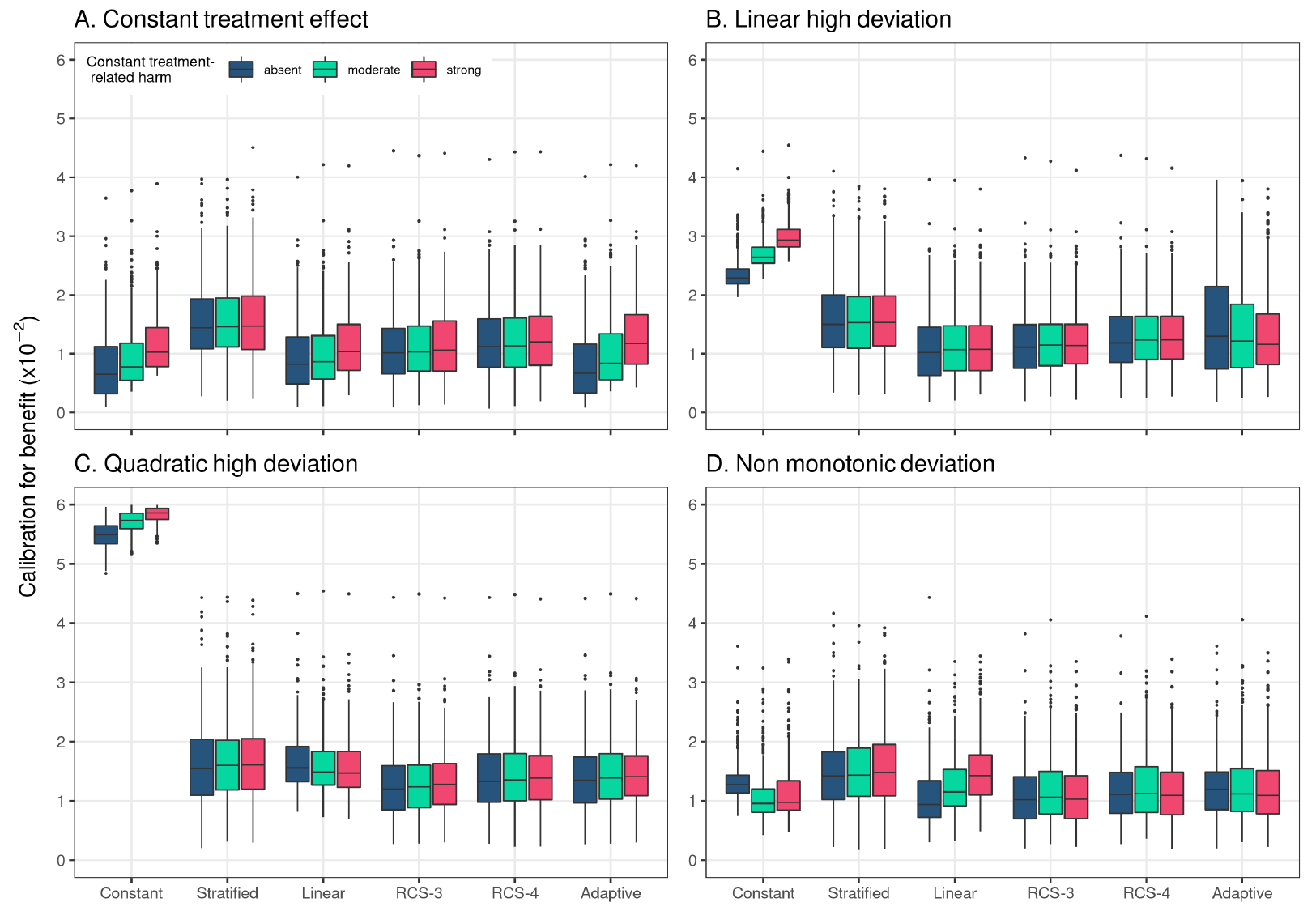} \caption{Calibration for benefit of the considered methods across 500 replications calculated in a simulated sample of size 500,000. True prediction AUC of 0.85 and sample size of 4,250}\label{fig:calibrationAuc}
\end{figure}

\newpage

\hypertarget{strong-relative-treatment-effect}{%
\section{Strong relative treatment effect}\label{strong-relative-treatment-effect}}

Here we present the root mean squared error of the considered methods using
strong constant relative treatment effect (\(\text{OR}=0.5\)) as the
reference. Again, the same sample size and prediction performance settings were
considered along with the same settings for linear, quadratic and non-monotonic
deviations from the base case scenario of constant relative treatment effects
are considered. All results can be found at
\url{https://arekkas.shinyapps.io/simulation_viewer/}.

\begin{figure}
\includegraphics[width=1\linewidth]{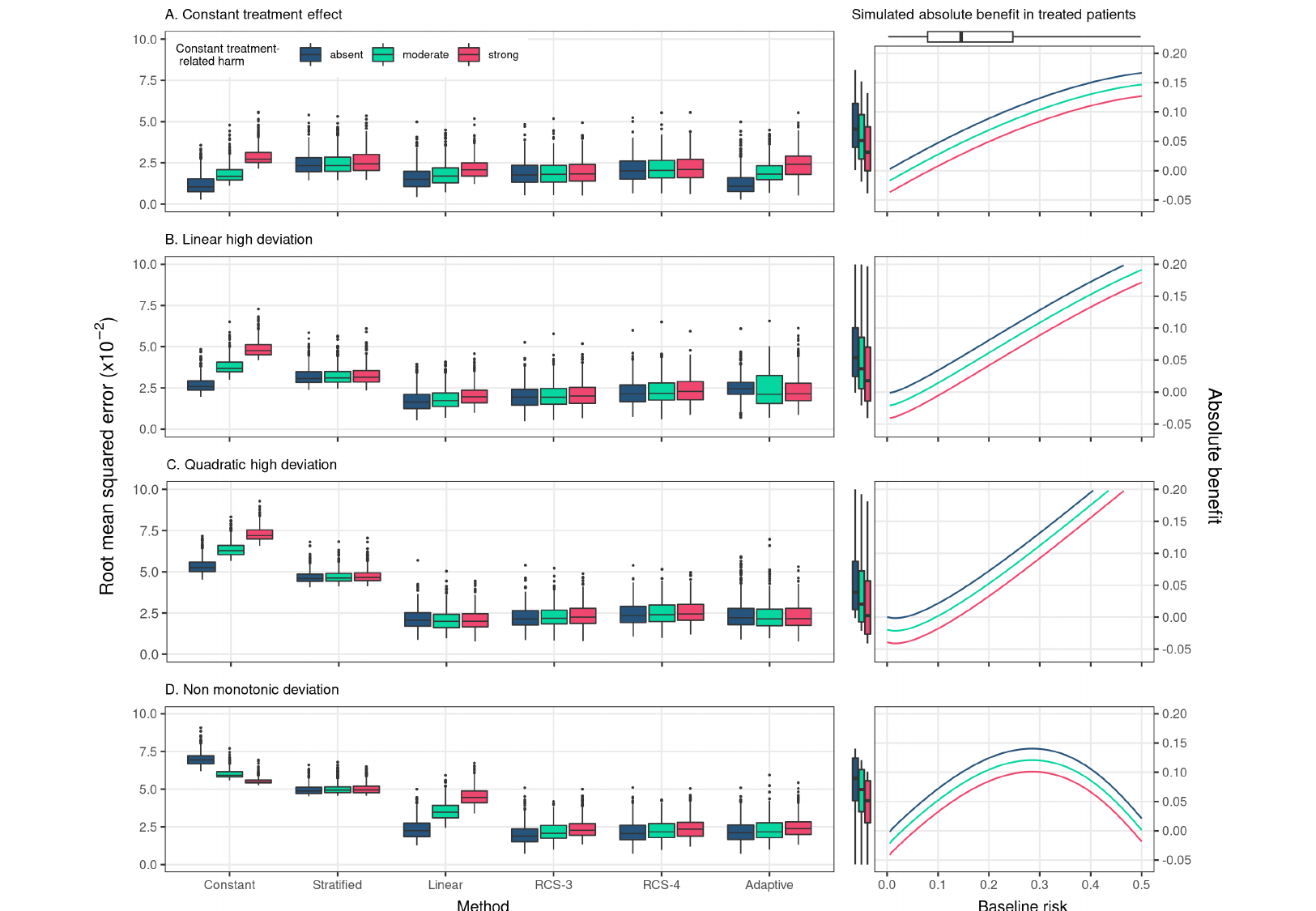} \caption{RMSE of the considered methods across 500 replications calculated in a simulated super-population of size 500,000. The scenario with true constant relative treatment effect (panel A) had a true prediction AUC of 0.75 and sample size of 4,250. The RMSE is also presented for strong linear (panel B), strong quadratic (panel C), and non-monotonic (panel D) deviations from constant relative treatment effects. Panels on the right side present the true relationship between baseline risk (x-axis) and absolute treatment benefit (y-axis). The 2.5, 25, 75 and 97.5 percentiles of the risk distribution are expressed in the boxplot.}\label{fig:rmseHighBase}
\end{figure}

\begin{figure}
\includegraphics[width=1\linewidth]{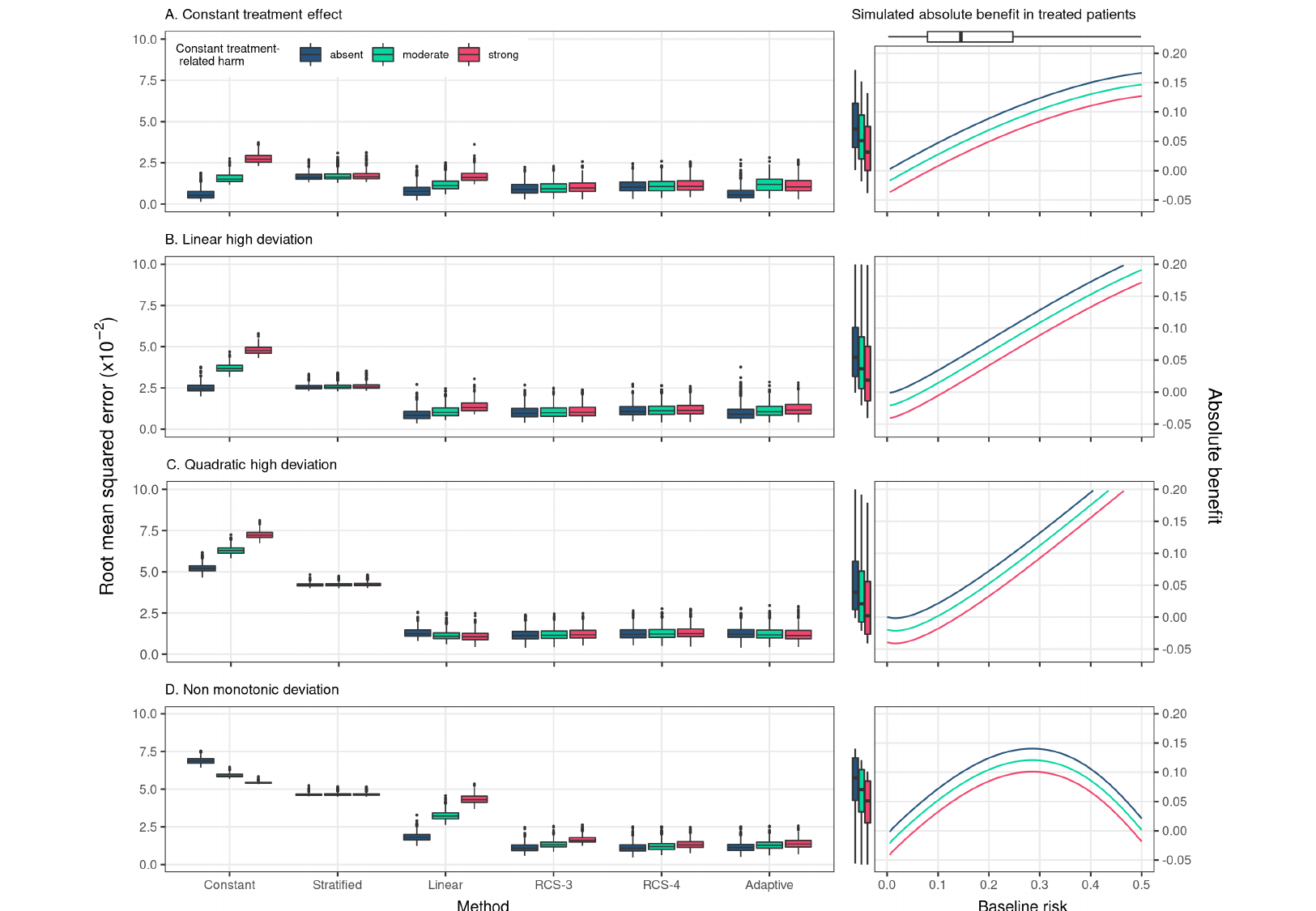} \caption{RMSE of the considered methods across 500 replcations calculated in a simulated sample of size 500,000. Sample size is 17,000 rather than 4,250 in Figure \ref{fig:rmseHighBase}.}\label{fig:rmseHighSampleSize}
\end{figure}

\begin{figure}
\includegraphics[width=1\linewidth]{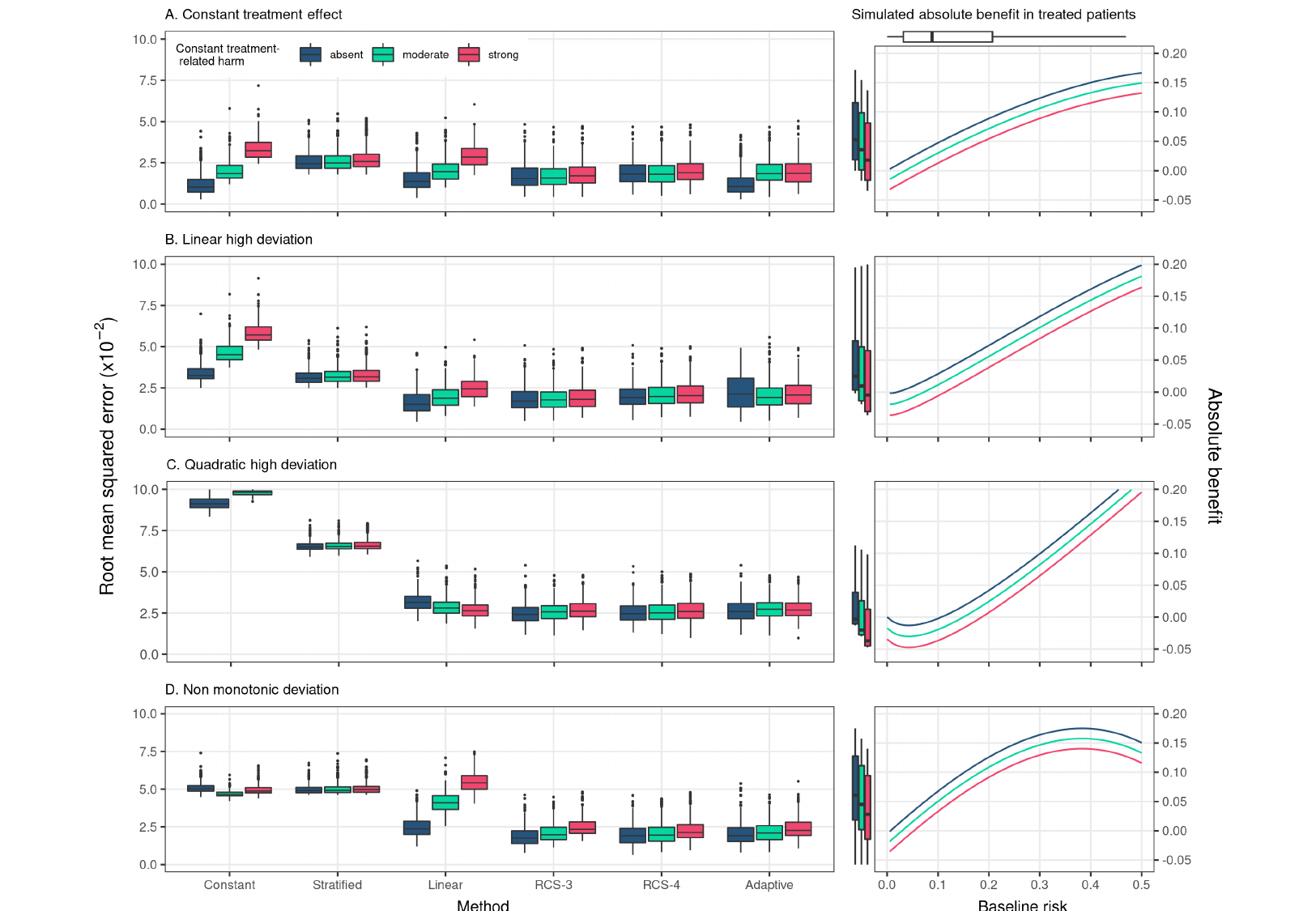} \caption{RMSE of the considered methods across 500 replcations calculated in a simulated sample of size 500,000. AUC is 0.85 rather than in Figure \ref{fig:rmseHighBase}.}\label{fig:rmseHighAuc}
\end{figure}

\newpage

\hypertarget{treatment-interactions}{%
\section{Treatment interactions}\label{treatment-interactions}}

We carried out a smaller set of simulations, in which we assumed true
treatment-covariate interactions. Sample size was set to 4,250 and the AUC of
the true prediction model was set to 0.75. The following scenarios were
considered: 1) 4 true weak positive interactions
(\(\text{OR}_{Z=1} / \text{OR}_{Z=0}=0.83\)); 2) 4 strong positive
interactions (\(\text{OR}_{Z=1} / \text{OR}_{Z=0}=0.61\)); 3) 2 weak and 2
strong positive interactions; 4) 4 weak negative interactions
(\(\text{OR}_{Z=1} / \text{OR}_{Z=0}=1.17\)); 5) 4 strong negative interactions
(\(\text{OR}_{Z=1} / \text{OR}_{Z=0}=1.39\)); 6) 2 weak and 2 strong negative
interactions; 7) combined positive and negative strong interactions. We also
considered constant treatment-related harms applied on the absolute scale to all
treated patients. The exact settings were: 1) absent treatment-related harms; 2)
moderate treatment-related harms, defined as 25\% of the average true benefit of the
scenario without treatment-related harms; 3) strong treatment-related harms
defined as 50\% of the true average benefit of the scenario without
treatment-related harms; 4) negative treatment-related harms (benefit), defined
as an absolute risk reduction for treated patients of 50\% of the true average
benefit of the scenario without treatment-related harms. The exact settings can
be found in Table \ref{tab:tab2}.

\begingroup\fontsize{6}{8}\selectfont

\begin{landscape}
\begin{longtable}[t]{rrrrrrrrrrrrrrrrrrrrrr}
\caption{\label{tab:tab2}Scenario settings of the treatment-covariate interaction scenarios.}\\
\toprule
\multicolumn{5}{c}{Analysis ID} & \multicolumn{9}{c}{Baseline risk} & \multicolumn{5}{c}{True treatment effect} & \multicolumn{2}{c}{Benefit} \\
\cmidrule(l{3pt}r{3pt}){1-5} \cmidrule(l{3pt}r{3pt}){6-14} \cmidrule(l{3pt}r{3pt}){15-19} \cmidrule(l{3pt}r{3pt}){20-21}
Scenario & Base & Type & N & AUC & Treatment-related harm & b0 & b1 & b2 & b3 & b4 & b5 & b6 & b7 & b8 & g0 & g1 & g2 & g5 & g6 & Before harms & After harms\\
\midrule
649 & interaction & weak & 4,250 & 0.75 & absent & -2.08 & 0.49 & 0.49 & 0.49 & 0.49 & 0.49 & 0.49 & 0.49 & 0.49 & -0.69 & -0.19 & -0.19 & -0.19 & -0.19 & 0.10 & 0.10\\
650 & interaction & weak & 4,250 & 0.75 & moderate-positive & -2.08 & 0.49 & 0.49 & 0.49 & 0.49 & 0.49 & 0.49 & 0.49 & 0.49 & -0.69 & -0.19 & -0.19 & -0.19 & -0.19 & 0.10 & 0.07\\
651 & interaction & weak & 4,250 & 0.75 & strong-positive & -2.08 & 0.49 & 0.49 & 0.49 & 0.49 & 0.49 & 0.49 & 0.49 & 0.49 & -0.69 & -0.19 & -0.19 & -0.19 & -0.19 & 0.10 & 0.05\\
652 & interaction & weak & 4,250 & 0.75 & negative & -2.08 & 0.49 & 0.49 & 0.49 & 0.49 & 0.49 & 0.49 & 0.49 & 0.49 & -0.69 & -0.19 & -0.19 & -0.19 & -0.19 & 0.10 & 0.12\\
\addlinespace
653 & interaction & mixed & 4,250 & 0.75 & absent & -2.08 & 0.49 & 0.49 & 0.49 & 0.49 & 0.49 & 0.49 & 0.49 & 0.49 & -0.69 & -0.19 & -0.49 & -0.19 & -0.49 & 0.10 & 0.10\\
654 & interaction & mixed & 4,250 & 0.75 & moderate-positive & -2.08 & 0.49 & 0.49 & 0.49 & 0.49 & 0.49 & 0.49 & 0.49 & 0.49 & -0.69 & -0.19 & -0.49 & -0.19 & -0.49 & 0.10 & 0.08\\
655 & interaction & mixed & 4,250 & 0.75 & strong-positive & -2.08 & 0.49 & 0.49 & 0.49 & 0.49 & 0.49 & 0.49 & 0.49 & 0.49 & -0.69 & -0.19 & -0.49 & -0.19 & -0.49 & 0.10 & 0.05\\
656 & interaction & mixed & 4,250 & 0.75 & negative & -2.08 & 0.49 & 0.49 & 0.49 & 0.49 & 0.49 & 0.49 & 0.49 & 0.49 & -0.69 & -0.19 & -0.49 & -0.19 & -0.49 & 0.10 & 0.13\\
\addlinespace
657 & interaction & strong & 4,250 & 0.75 & absent & -2.08 & 0.49 & 0.49 & 0.49 & 0.49 & 0.49 & 0.49 & 0.49 & 0.49 & -0.69 & -0.49 & -0.49 & -0.49 & -0.49 & 0.11 & 0.11\\
658 & interaction & strong & 4,250 & 0.75 & moderate-positive & -2.08 & 0.49 & 0.49 & 0.49 & 0.49 & 0.49 & 0.49 & 0.49 & 0.49 & -0.69 & -0.49 & -0.49 & -0.49 & -0.49 & 0.11 & 0.08\\
659 & interaction & strong & 4,250 & 0.75 & strong-positive & -2.08 & 0.49 & 0.49 & 0.49 & 0.49 & 0.49 & 0.49 & 0.49 & 0.49 & -0.69 & -0.49 & -0.49 & -0.49 & -0.49 & 0.11 & 0.06\\
660 & interaction & strong & 4,250 & 0.75 & negative & -2.08 & 0.49 & 0.49 & 0.49 & 0.49 & 0.49 & 0.49 & 0.49 & 0.49 & -0.69 & -0.49 & -0.49 & -0.49 & -0.49 & 0.11 & 0.14\\
\addlinespace
661 & interaction & negative-weak & 4,250 & 0.75 & absent & -2.08 & 0.49 & 0.49 & 0.49 & 0.49 & 0.49 & 0.49 & 0.49 & 0.49 & -0.69 & 0.16 & 0.16 & 0.16 & 0.16 & 0.06 & 0.06\\
662 & interaction & negative-weak & 4,250 & 0.75 & moderate-positive & -2.08 & 0.49 & 0.49 & 0.49 & 0.49 & 0.49 & 0.49 & 0.49 & 0.49 & -0.69 & 0.16 & 0.16 & 0.16 & 0.16 & 0.06 & 0.05\\
663 & interaction & negative-weak & 4,250 & 0.75 & strong-positive & -2.08 & 0.49 & 0.49 & 0.49 & 0.49 & 0.49 & 0.49 & 0.49 & 0.49 & -0.69 & 0.16 & 0.16 & 0.16 & 0.16 & 0.06 & 0.03\\
664 & interaction & negative-weak & 4,250 & 0.75 & negative & -2.08 & 0.49 & 0.49 & 0.49 & 0.49 & 0.49 & 0.49 & 0.49 & 0.49 & -0.69 & 0.16 & 0.16 & 0.16 & 0.16 & 0.06 & 0.08\\
\addlinespace
665 & interaction & negative-mixed & 4,250 & 0.75 & absent & -2.08 & 0.49 & 0.49 & 0.49 & 0.49 & 0.49 & 0.49 & 0.49 & 0.49 & -0.69 & 0.16 & 0.33 & 0.16 & 0.33 & 0.05 & 0.05\\
666 & interaction & negative-mixed & 4,250 & 0.75 & moderate-positive & -2.08 & 0.49 & 0.49 & 0.49 & 0.49 & 0.49 & 0.49 & 0.49 & 0.49 & -0.69 & 0.16 & 0.33 & 0.16 & 0.33 & 0.05 & 0.04\\
667 & interaction & negative-mixed & 4,250 & 0.75 & strong-positive & -2.08 & 0.49 & 0.49 & 0.49 & 0.49 & 0.49 & 0.49 & 0.49 & 0.49 & -0.69 & 0.16 & 0.33 & 0.16 & 0.33 & 0.05 & 0.03\\
668 & interaction & negative-mixed & 4,250 & 0.75 & negative & -2.08 & 0.49 & 0.49 & 0.49 & 0.49 & 0.49 & 0.49 & 0.49 & 0.49 & -0.69 & 0.16 & 0.33 & 0.16 & 0.33 & 0.05 & 0.06\\
\addlinespace
669 & interaction & negative-strong & 4,250 & 0.75 & absent & -2.08 & 0.49 & 0.49 & 0.49 & 0.49 & 0.49 & 0.49 & 0.49 & 0.49 & -0.69 & 0.33 & 0.33 & 0.33 & 0.33 & 0.04 & 0.04\\
670 & interaction & negative-strong & 4,250 & 0.75 & moderate-positive & -2.08 & 0.49 & 0.49 & 0.49 & 0.49 & 0.49 & 0.49 & 0.49 & 0.49 & -0.69 & 0.33 & 0.33 & 0.33 & 0.33 & 0.04 & 0.03\\
671 & interaction & negative-strong & 4,250 & 0.75 & strong-positive & -2.08 & 0.49 & 0.49 & 0.49 & 0.49 & 0.49 & 0.49 & 0.49 & 0.49 & -0.69 & 0.33 & 0.33 & 0.33 & 0.33 & 0.04 & 0.02\\
672 & interaction & negative-strong & 4,250 & 0.75 & negative & -2.08 & 0.49 & 0.49 & 0.49 & 0.49 & 0.49 & 0.49 & 0.49 & 0.49 & -0.69 & 0.33 & 0.33 & 0.33 & 0.33 & 0.04 & 0.05\\
\addlinespace
673 & interaction & combined & 4,250 & 0.75 & absent & -2.08 & 0.49 & 0.49 & 0.49 & 0.49 & 0.49 & 0.49 & 0.49 & 0.49 & -0.69 & -0.49 & 0.33 & -0.49 & 0.33 & 0.08 & 0.08\\
674 & interaction & combined & 4,250 & 0.75 & moderate-positive & -2.08 & 0.49 & 0.49 & 0.49 & 0.49 & 0.49 & 0.49 & 0.49 & 0.49 & -0.69 & -0.49 & 0.33 & -0.49 & 0.33 & 0.08 & 0.06\\
675 & interaction & combined & 4,250 & 0.75 & strong-positive & -2.08 & 0.49 & 0.49 & 0.49 & 0.49 & 0.49 & 0.49 & 0.49 & 0.49 & -0.69 & -0.49 & 0.33 & -0.49 & 0.33 & 0.08 & 0.04\\
676 & interaction & combined & 4,250 & 0.75 & negative & -2.08 & 0.49 & 0.49 & 0.49 & 0.49 & 0.49 & 0.49 & 0.49 & 0.49 & -0.69 & -0.49 & 0.33 & -0.49 & 0.33 & 0.08 & 0.10\\
\bottomrule
\end{longtable}
\end{landscape}
\endgroup{}

\begin{figure}
\includegraphics[width=1\linewidth]{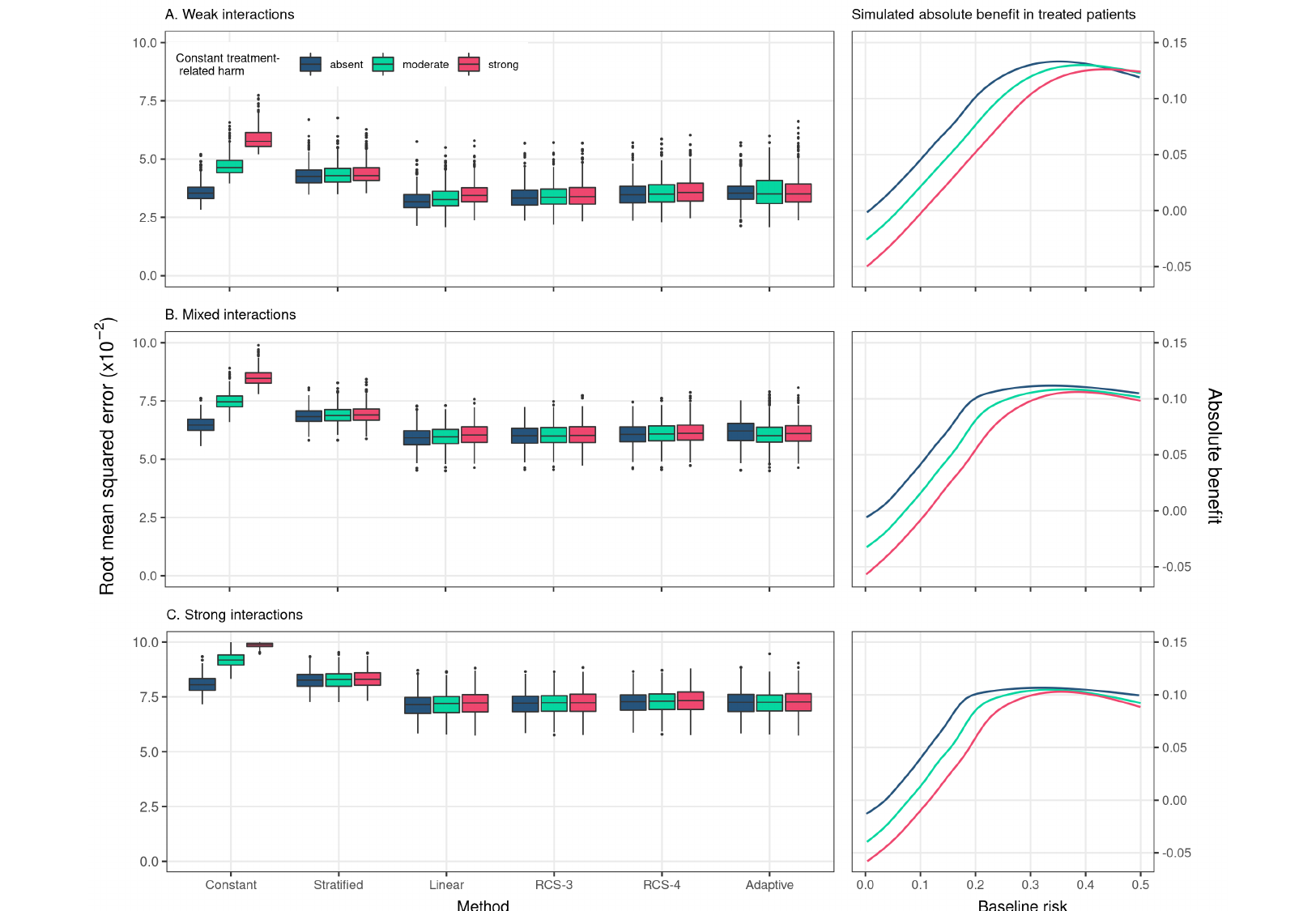} \caption{RMSE of the considered methods across 500 replications calculated in a simulated sample of size 500,000 where treatment-covariate interactions all favoring treatment were considered.}\label{fig:rmseInteractionPositive}
\end{figure}
\begin{figure}
\includegraphics[width=1\linewidth]{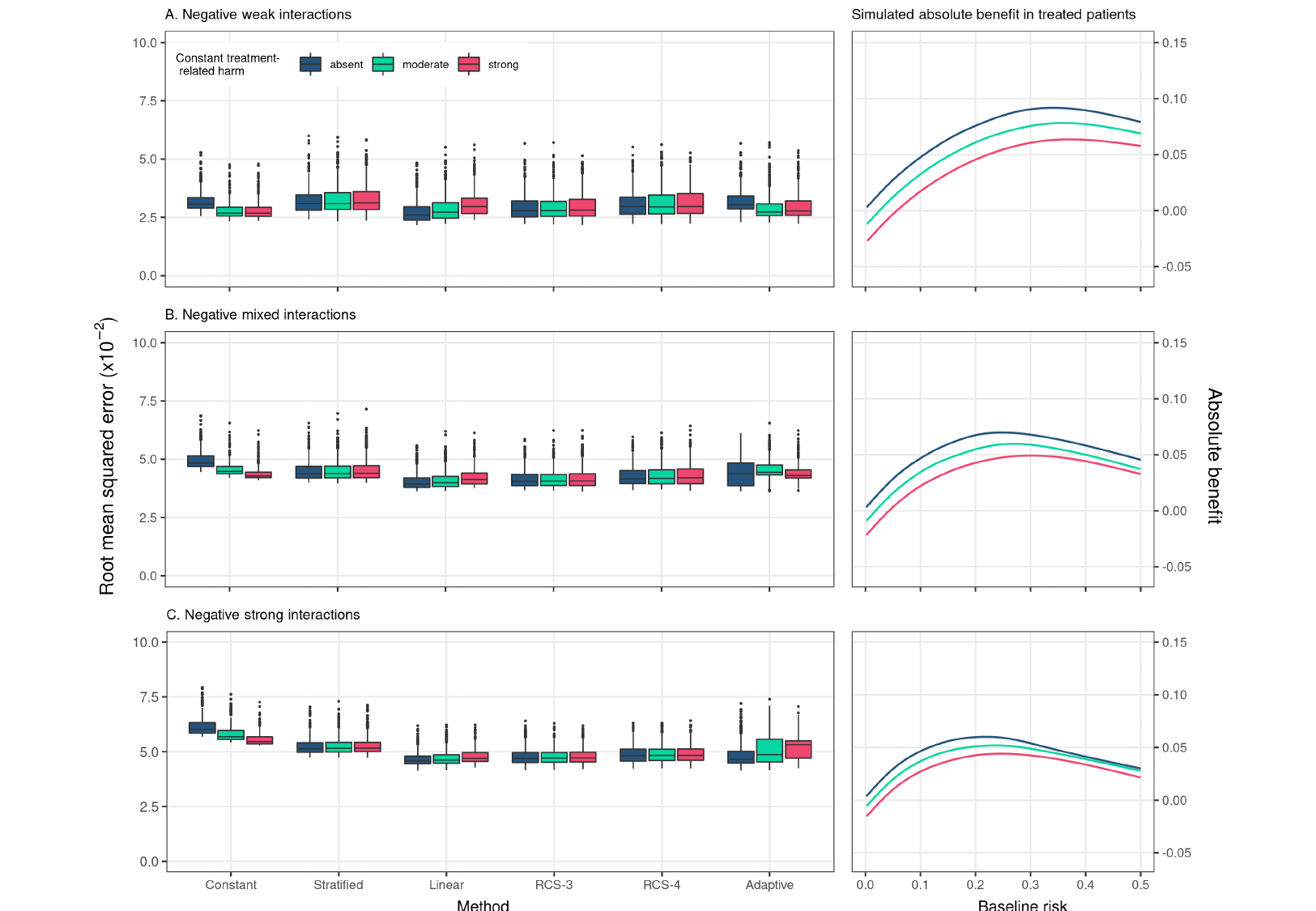} \caption{RMSE of the considered methods across 500 replications calculated in a simulated sample of size 500,000 where treatment-covariate interactions all favoring the control were considered.}\label{fig:rmseInteractionNegative}
\end{figure}
\begin{figure}
\includegraphics[width=1\linewidth]{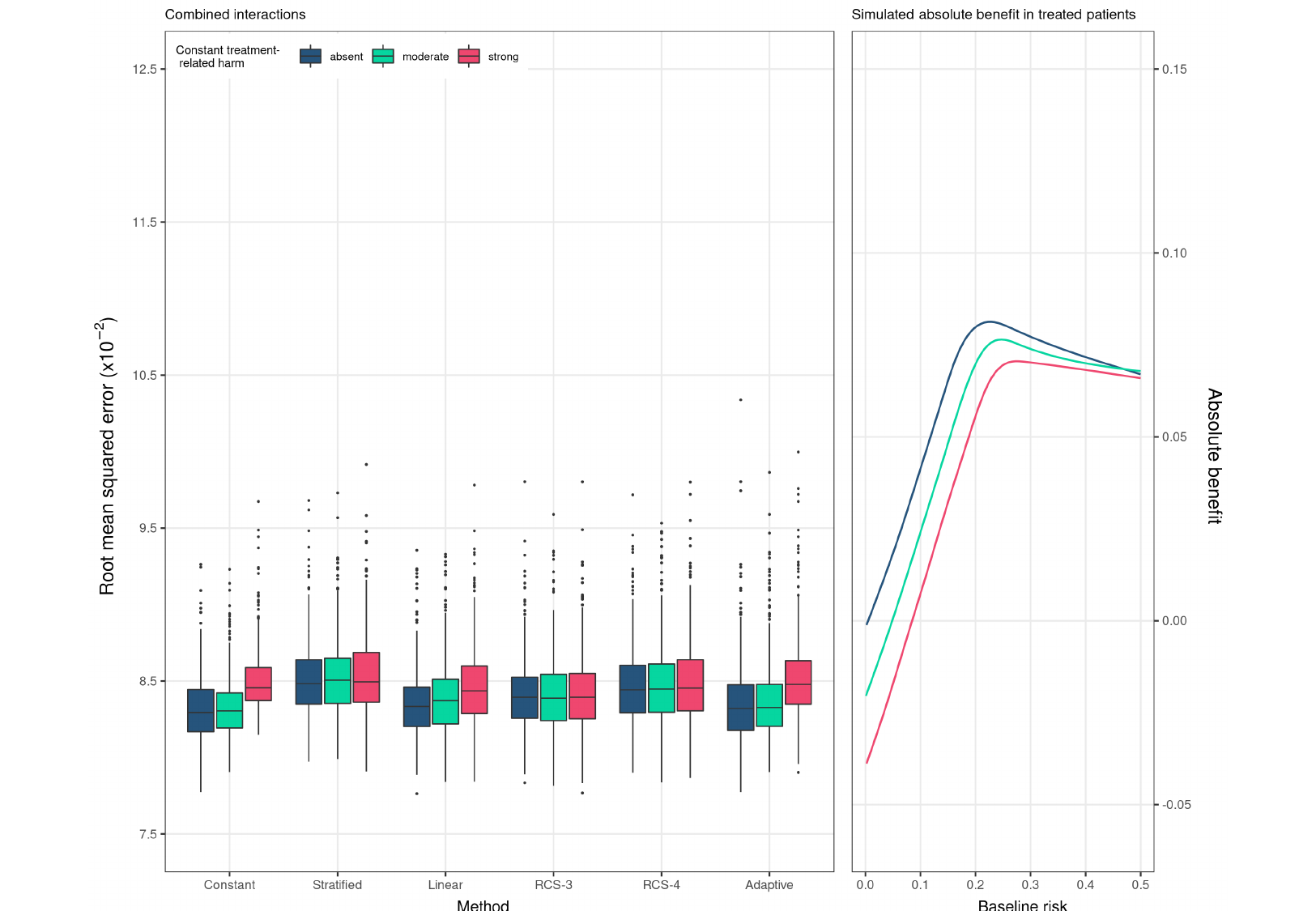} \caption{RMSE of the considered methods across 500 replications calculated in a simulated sample of size 500,000 where treatment-covariate interactions 2 favoring treatment and 2 favoring the control were considered.}\label{fig:rmseInteractionCombined}
\end{figure}

\newpage

\hypertarget{empirical-illustration}{%
\section{Empirical illustration}\label{empirical-illustration}}

For predicting baseline risk of 30-day mortality we fitted a logistic regression
model with age, Killip class (\emph{Killip}), systolic blood pressure (\emph{sysbp}),
pulse rate (\emph{pulse}), prior myocardial infarction (\emph{pmi}), location of
myocardial infarction (\emph{miloc}) and treatment as the covariates. Baseline
predictions were made setting treatment to 0.

\begin{equation*}
P(outcome=1|X=x) = \text{expit}(lp(x)),
\label{eq:gusto1}
\end{equation*}
where
\begin{equation*}
\begin{aligned}
lp(x)=& \beta_0 + \beta_1 \text{age} + \beta_2 I(\text{Killip}=II) + \beta_3I(\text{Killip}=III) +\\
&\beta_4 I(\text{Killip}=IV) + \beta_5min(\text{sysbp}, 120) + \beta_6 \text{pulse}+\\
&\beta_7 max(\text{pulse - 50, 0}) + \beta_8 I(\text{pmi}=yes)+\\
&\beta_9 I(\text{miloc}=Anterior) + \beta_9 I(\text{miloc}=Other) +\\
&\gamma\times\text{treatment}
\end{aligned}
\label{eq:gusto2}
\end{equation*}

and \(expit(x) = \frac{e^x}{1+e^x}\)

\begingroup\fontsize{7}{9}\selectfont

\begin{longtable}[t]{lrrrr}
\caption{\label{tab:unnamed-chunk-5}Coefficients of the prediction model for 30-day mortality, based on the data from GUSTO-I trial.}\\
\toprule
Variable & Estimate & stderror & zvalue & pvalue\\
\midrule
Intercept & -3.020 & 0.797 & -3.788 & 0.000\\
Age & -0.208 & 0.053 & -3.935 & 0.000\\
Killip class = II & 0.077 & 0.002 & 31.280 & 0.000\\
Killip class = III & 0.614 & 0.059 & 10.423 & 0.000\\
Killip class = IV & 1.161 & 0.121 & 9.566 & 0.000\\
Systolic blood pressure & 1.921 & 0.162 & 11.872 & 0.000\\
Pulse rate (1) & -0.039 & 0.002 & -20.332 & 0.000\\
Pulse rate (2) & -0.024 & 0.016 & -1.521 & 0.128\\
Previous MI (yes) & 0.043 & 0.016 & 2.675 & 0.007\\
MI location (Other) & 0.447 & 0.056 & 7.964 & 0.000\\
MI location (Anterior) & 0.286 & 0.135 & 2.126 & 0.033\\
Treatment & 0.543 & 0.051 & 10.625 & 0.000\\
\bottomrule
\end{longtable}
\endgroup{}

\newpage

\hypertarget{references}{%
\section{References}\label{references}}

\setlength{\parindent}{-0.25in}
\setlength{\leftskip}{0.25in}

\noindent

\hypertarget{refs}{}
\begin{CSLReferences}{0}{0}
\leavevmode\vadjust pre{\hypertarget{ref-Kent2016}{}}%
\CSLLeftMargin{{[}1{]} }%
\CSLRightInline{Kent DM, Nelson J, Dahabreh IJ, Rothwell PM, Altman DG, Hayward RA. Risk and treatment effect heterogeneity: Re-analysis of individual participant data from 32 large clinical trials. International Journal of Epidemiology 2016:dyw118. \url{https://doi.org/10.1093/ije/dyw118}.}

\leavevmode\vadjust pre{\hypertarget{ref-harrell2017regression}{}}%
\CSLLeftMargin{{[}2{]} }%
\CSLRightInline{Harrell FE. Regression modeling strategies. vol. 330. Springer; 2017.}

\leavevmode\vadjust pre{\hypertarget{ref-Austin2019}{}}%
\CSLLeftMargin{{[}3{]} }%
\CSLRightInline{Austin PC, Steyerberg EW. The integrated calibration index ({ICI}) and related metrics for quantifying the calibration of logistic regression models. Statistics in Medicine 2019;38:4051--65. \url{https://doi.org/10.1002/sim.8281}.}

\end{CSLReferences}

\setlength{\parindent}{0in}
\setlength{\leftskip}{0in}

\noindent